\documentclass[aps,prd,preprint,superscriptaddress,preprintnumbers,floatfix]{revtex4-1}

\pdfoutput=1
\usepackage{amsmath,amssymb,bm,bbm,mathrsfs,amsfonts}
\usepackage[colorlinks=true,linkcolor=blue,citecolor=blue,urlcolor=blue]{hyperref}
\usepackage{graphicx,booktabs,tabularx,multirow}
\usepackage[dvipsnames]{xcolor}
\usepackage{colortbl}
\usepackage{soul}

\definecolor{TableLightOrange}{RGB}{255,229,204}
\definecolor{TableLightRed}{RGB}{255,204,204}

\newcommand{\dd}[1]{\mathrm{d}#1}

\newcommand{\xpi}{x_\pi}

\begin{document}
\preprint{JLAB-THY-26-4993}

\title{Constraints on pion parton distributions at small $x$ \\ from leading-neutron electroproduction data}

\author{C. Yang}
\affiliation{The University of Texas at Austin, Austin, Texas 78705, USA}
\affiliation{Temple University, Philadelphia, Pennsylvania 19122, USA}
\author{P. C. Barry}
\affiliation{Physics Division, Argonne National Laboratory, Lemont, Illinois 60439, USA}
\affiliation{Jefferson Lab, Newport News, Virginia 23606, USA\\
        \vspace*{0.2cm}
        {\bf JAM Collaboration \\ \footnotesize{~(Pion PDF Analysis Group)}}}
\author{W. Melnitchouk}
\author{N. Sato}
\affiliation{Jefferson Lab, Newport News, Virginia 23606, USA\\
        \vspace*{0.2cm}
        {\bf JAM Collaboration \\ \footnotesize{~(Pion PDF Analysis Group)}}}

\begin{abstract}
We revisit the extraction of parton distribution functions (PDFs) in the pion in a global QCD analysis of pion--nucleus Drell--Yan and leading-neutron (LN) electroproduction data, in view of recent suggestions that uncertainties on the pion PDFs at small parton momentum fraction $x$ are dramatically underestimated.
We perform a set of controlled fits that vary the LN data coverage, PDF parametrization, and positivity constraints, determining both the PDFs and pion-nucleon splitting functions simultaneously.
While the use of a restricted H1 LN data sample used by the Fant\^omas group is found to enlarge the PDF uncertainties and admit solutions with vanishing gluon PDFs, our global analysis provides a good description of an extended H1 and ZEUS LN database, significantly reducing the parametrization dependence and strongly disfavoring a zero gluon distribution.
\end{abstract}

\maketitle

\section{Introduction}
\label{sec:intro}

As the lightest QCD bound state, the pion has long been known to play a central role in hadron structure and the dynamics of the strong nuclear interactions~\cite{Ericson:1988gk}. 
Its quark and gluon momentum distributions, on the other hand, remain poorly determined compared with those of the proton, primarily due to the absence of stable pion targets.
Existing phenomenological analyses rely predominantly on pion--nucleus Drell--Yan (DY) lepton-pair production data~\cite{NA10:1985ibr, Conway:1989fs}, which constrain the valence distribution at moderate and large parton momentum fractions $x$, and semi-inclusive leading-neutron (LN) electroproduction data on protons from HERA~\cite{ZEUS:2002gig, H1:2010hym}.
The latter provide indirect access to nearly on-shell pions at smaller $x$ values, and in principle allow discrimination between the valence, sea quark, and gluon distributions~\cite{McKenney:2015xis}.

In a series of studies, the JAM Collaboration used a Bayesian Monte Carlo framework to extract from these complementary datasets the pion PDFs across an extended range of~$x$, observing directly for the first time a substantial gluon momentum fraction in the pion at small~$x$~\cite{Barry:2018ort}. 
Subsequent studies incorporated transverse momentum dependent DY measurements at large $q_T$ to explore their impact on the gluon PDF in the pion~\cite{Cao:2021aci}, although the uncertainties on those data were too large to provide significant constraints. 
At large~$x$ the effects of threshold resummation were critically evaluated to determine the $x \to 1$ behavior of the valence quark PDF~\cite{Barry:2021osv}, and a later study combined for the first time the experimental and lattice QCD observables and quantified both pion PDF uncertainties and lattice systematic effects~\cite{Barry:2022aix}.

\begin{figure}[t]
\centering
\includegraphics[width=\linewidth]{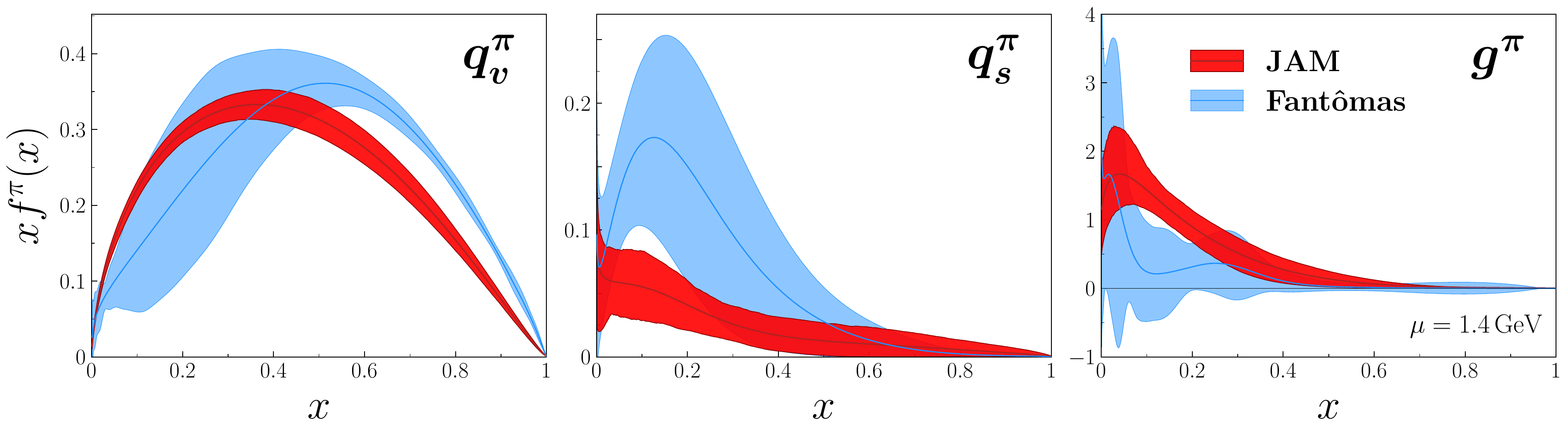}
\caption{Comparison of the previous JAM~\cite{Barry:2021osv} (red bands) and Fant\^{o}mas~\cite{Kotz:2023pbu} (blue bands) pion PDFs for the valence (left), sea quark (middle), and gluon (right) distributions at the scale $\mu=1.4$~GeV. The bands correspond to 68\% credible intervals.}
\label{fig:jam-fantomas-comparison}
\end{figure}

More recently, an analysis by the Fant\^{o}mas Collaboration~\cite{Kotz:2023pbu}, using a different treatment of the LN data and a more flexible B\'ezier-based PDF parametrization, reported substantially larger pion valence PDF uncertainties, along with fitted gluon distributions consistent with zero at the input scale.
The comparison between the JAM and Fant\^{o}mas results for the valence ($q_v^\pi$), sea-quark ($q_s^\pi$), and gluon ($g^\pi$) PDFs at $\mu=1.4~{\rm GeV}$ is illustrated in Fig.~\ref{fig:jam-fantomas-comparison}.
While the two analyses give broadly similar valence shapes, the Fant\^{o}mas uncertainty is substantially larger, especially at small $x$ values.
The contrast is more dramatic for the sea-quark and gluon distributions, where the Fant\^{o}mas bands are much broader, and the gluon is compatible with zero over a large part of the fitted $x$ range.

In addition to understanding the intrinsic partonic content of the pion, especially its breakdown into quark and gluon contributions, knowledge of the gluon PDF is essential for determining the gluonic contribution to the pion's energy-momentum tensor and its mass decomposition~\cite{Ji:1994av, Ji:1995sv, Metz:2020vxd}.
In view of the Fant\^{o}mas findings about the potential bias introduced in the earlier JAM analyses through the reliance on overly restricted parametric forms, we revisit the constraints from the LN data and explore various aspects of the data selection, PDF parametrization, and fitting methodology in order to test the robustness of the previous conclusions.

In particular, we explore in detail the extent to which the restricted LN data sample utilized in the Fant\^{o}mas analysis~\cite{Kotz:2023pbu}, based on a subset of H1 data~\cite{H1:2010hym}, weakens the constraints on the sea quark--gluon separation, and conversely the ability of the full LN dataset that includes also measurements from ZEUS~\cite{ZEUS:2002gig} to improve the determination of the gluon PDF and exclude zero-gluon solutions.
We further examine and contrast results using a more flexible, two-component quark parametrization with the traditional single-shape functional form, and identify the degree to which the uncertainty enhancement is associated with replicas in which the valence and sea-quark PDFs become negative, and the impact of positivity constraints on the uncertainty bands.
To reduce the model dependence of the analysis, we fit the pion--nucleon splitting functions to the HERA LN data rather than importing these from external sources~\cite{Kotz:2023pbu}, and perform scans in the neutron longitudinal momentum fraction to empirically determine the region of kinematics that can be described through the one-pion exchange mechanism.
As a consistency check, we further benchmark the splitting function results with data from inclusive neutron production in $pp$ scattering.

We begin in Sec.~\ref{sec:setup} by describing our theoretical framework, data selections, and PDF parametrizations.
In Sec.~\ref{sec:results} we present the controlled comparison with the Fant\^omas analysis, examining in particular the interplay between data coverage and parametrization flexibility.
The results of the JAM analysis including the full H1 and ZEUS LN datasets are presented in Sec.~\ref{sec:extensions}, for both the fitted splitting functions and the pion PDFs, and compared with previous JAM determinations.
Finally, we summarize our results in Sec.~\ref{sec:conclusion}.

\section{Analysis setup}
\label{sec:setup}

In this section, we describe the theoretical framework and analysis setup used to extract the pion PDFs.
We summarize the treatment of pion–nucleus DY and LN electroproduction, introduce the pion–nucleon splitting functions and regulator models, and discuss the experimental datasets used in the analysis.
We also discuss the PDF parametrizations, positivity constraints, and Bayesian Monte Carlo methodology employed in the fit comparisons.

\subsection{Theoretical framework}
\label{ssec:theory}

The experimental information on pion PDFs in this analysis comes from the complementary pion--nucleus DY lepton-pair production and LN electroproduction processes. 
Historically~\cite{Owens:1984zj, Aurenche:1989sx, Sutton:1991ay, Gluck:1991ey, Gluck:1999xe, Wijesooriya:2005ir, Aicher:2010cb, Novikov:2020snp}, the DY data have been known to primarily constrain the valence-quark PDF at moderate and large $x$ values, whereas the LN data extend the sensitivity to smaller $x$ and provide the principal constraint on the sea-quark and gluon distributions.

\subsubsection{Drell-Yan lepton-pair production}
\label{sssec:DY}

In pion-induced DY lepton-antilepton (usually muons) production, a parton from the incident pion annihilates with a parton from a nucleon in the target nucleus to produce a virtual photon which subsequently decays into a lepton pair, $\pi N \to \gamma^* X \to \mu^+ \mu^- X$.
In collinear factorization the cross section can be written as~\cite{Drell:1970wh, Altarelli:1979ub, Collins:1983ju}
\begin{align}
 \frac{\mathrm{d}^2\sigma}{\dd Q^2\,\dd Y}
 = & \frac{4\pi\alpha^2}{9 Q^2 S}
 \sum_{ij}
 \int_{x_\pi}^1\frac{\dd\hat x_\pi}{\hat x_\pi}
 \int_{x_N}^1\frac{\dd\hat x_N}{\hat x_N}
 C_{ij}
 \big(\hat x_\pi,\hat x_N,x_\pi,x_N,Q/\mu\big)\,
 f_i^\pi(\hat x_\pi,\mu)\,
 f_j^{N/A}(\hat x_N,\mu),
 \label{eq:dyfactorization}
\end{align}
where $S$ is the pion--bound nucleon center of mass energy squared, and $Q$ and $Y$ are the invariant mass and rapidity of the lepton pair.
The functions $f_i^\pi$ and $f_j^{N/A}$ are the pion and bound nucleon PDFs evaluated at parton momentum fractions $\hat x_\pi$ and $\hat x_N$, respectively, at a factorization scale $\mu$, and $C_{ij}$ are the perturbatively calculable hard scattering coefficients. 
For the DY process we include threshold resummation of the hard coefficients at next-to-leading-logarithmic accuracy using the double Mellin method, following the implementation of Refs.~\cite{Barry:2021osv,Westmark:2017uig}.  
This procedure resums the logarithmically enhanced soft-gluon contributions that become important near partonic threshold.

The hadronic momentum fractions $x_{\pi,N}$ satisfy $x_{\pi,N} = \sqrt{\tau} e^{\pm Y}$, with $\tau=Q^2/S$, and at leading order coincide with the corresponding partonic fractions, $\hat x_{\pi,N}$.
The E615 and NA10 cross sections are commonly presented in terms of the Feynman scaling variable, $x_F = x_\pi - x_N$.
Following previous JAM pion analyses~\cite{Barry:2018ort, Cao:2021aci, Barry:2021osv, Barry:2022aix}, the same DY dataset and theory treatment are held fixed in all fits considered here, so that any differences found between fit scenarios arise from the LN selection and PDF assumptions.

\subsubsection{Leading neutron electroproduction}
\label{sssec:LN}

In LN electroproduction, an electron (four-momentum $\ell$) scatters from a proton (four-momentum $p$) to a final state electron ($\ell'$) and recoil neutron ($p'$), together with undetected hadronic debris $X$, with the neutron produced in the forward direction carrying a fraction $x_L$ of the proton's longitudinal momentum or energy,
\begin{equation}
x_L\, =\, \frac{p' \cdot \ell}{p \cdot \ell}\, \approx\, \frac{p'_0}{p_0}.
\label{eq:xL}
\end{equation}
At forward scattering angles, the semi-inclusive reaction is dominated by the $t$-channel process, in which the exchanged virtual photon (four-momentum $q = \ell - \ell'$) scatters from a virtual pion (four-momentum $k = p - p'$) carrying a fraction 
\begin{equation}
\bar{x}_L = 1 - x_L
\label{eq:xLbar}
\end{equation}
of the proton's energy.
The four-momentum transfer squared between the initial proton and final state neutron is given by
\begin{equation}
t \equiv k^2 
  = -\frac{1}{x_L} \big( k_T^2 + \bar{x}_L^2 M^2 \big),
\label{eq:t}
\end{equation}
where $k_T$ is the transverse momentum of the exchanged pion (which in the target rest frame or center of mass frame also corresponds to the magnitude of the transverse momentum of the produced neutron) and $M$ is the nucleon mass. 
In the region where the pion is not far from its mass shell, the semi-inclusive LN structure function can be written in factorized form as~\cite{Sullivan:1971kd}
\begin{equation}
F_2^{\rm LN (4)}(x_B,Q^2,\bar{x}_L,k_T^2)\,
\approx\, 2 \tilde{f}_{\pi N}(\bar{x}_L, k_T^2)\, F_2^\pi(\xpi,Q^2),
\label{eq:lnfactorization}
\end{equation}
where $x_B = Q^2 / 2 p \cdot q$ is the Bjorken scaling variable for the LN process, $\tilde{f}_{\pi N}(\bar{x}_L, k_T^2)$ is the unintegrated $p \to \pi N$ splitting function (sometimes referred to as the ``pion flux''), and $F_2^\pi(\xpi,Q^2)$ is the inclusive DIS pion structure function, with
\begin{equation}
\xpi = \frac{x_B}{\,\bar{x}_L}\, .
\label{eq:xdefs}
\end{equation}
Note that the factor 2 on the right-hand-side of Eq.~(\ref{eq:lnfactorization}) arises from the isospin Clebsch-Gordan coefficient for the $p \to \pi^+ n$ transition relevant for the LN data.
The validity of Eq.~\eqref{eq:lnfactorization} is expected to be strongest in the forward-neutron region, $\bar x_L \to 0$, and motivates the kinematic scans discussed below.

Within collinear factorization, the pion structure function can be written as a convolution of perturbative coefficient functions and pion quark, antiquark, and gluon PDFs,
\begin{equation}
 F_2^\pi(x_\pi,Q^2)
 = x_\pi \sum_q e_q^2
 \Big[C_{q} \otimes (q^\pi+\bar q^\pi) + C_{g}\otimes g^\pi\Big](\xpi,Q^2),
 \label{eq:f2pifactorization}
\end{equation}
where $C_{q}$ and $C_{g}$ are the DIS hard scattering coefficient functions,
$e_q$ is the quark electric charge, and $\otimes$ denotes the convolution integral,
\begin{equation}
\big[C\otimes f\big](x) \equiv \int_{x}^1 \frac{\dd z}{z}\, C(z) \, f\Big(\frac{x}{z}\Big). 
\label{eq:conv}
\end{equation}
Note that $x_\pi$ is an experimentally measured quantity, and not the momentum fraction of partons within the pion. 
At leading order in the strong coupling $\alpha_s$, $C_{q}(z) \to \delta(1-z)$ and $C_{g}(z) \to 0$, so that $F_2^\pi = x_\pi \sum_q e_q^2\, (q^\pi+\bar q^\pi)$, as in the quark-parton model. 
The gluon PDF contributes at higher orders in $\alpha_s$ and through the $Q^2$ evolution of the quark distributions. 
The broad $Q^2$ lever arm of the LN data therefore provides sensitivity to the pion gluon PDF through scaling violations.

For $\bar x_L > 0$, the unintegrated $\pi N$ splitting function is given by~\cite{Sullivan:1971kd, Thomas:1983fh, Melnitchouk:1992yd, Burkardt:2012hk, Salamu:2014pka},
\begin{eqnarray}
\tilde{f}_{\pi N}(\bar x_L, k_T^2)
&=& \frac{g_A^2M^2}{(4\pi f_\pi)^2} 
  \frac{\bar x_L \big( k_T^2+\bar x_L^2M^2 \big)}{x_L^2\, D_{\pi N}^2}\,
  \big|\mathcal{F}\big|^2,
\label{eq:splitting_function}
\end{eqnarray}
where 
\begin{equation}
D_{\pi N}\ \equiv\ t-m_\pi^2\
=\ - \frac{k_T^2 + \bar x_L^2 M^2 + x_L m_\pi^2}{x_L}
\label{eq:DpiN}
\end{equation}
is the pion propagator, $m_\pi$ is the pion mass, $g_A$ is the axial charge, $f_\pi$ is the pion decay constant, and $\mathcal F$ is an ultraviolet regulator. 
To examine the model dependence of our fits, we consider a family of models for the regulator $\mathcal F$~\cite{Zoller:1991cb, Melnitchouk:1992yd, Holtmann:1996be, McKenney:2015xis},
\begin{equation}
{\mathcal F} = \left\{
\begin{array}{lcll}
&\hspace*{-0.35cm} &
 \hspace*{-0.05cm}\exp\big[ (M^2-s)/\Lambda^2 \big]
	&\quad s\text{-dependent exponential},		\\
&\hspace*{-0.35cm} &
 \hspace*{-0.05cm}\exp\big[ D_{\pi N}/\Lambda^2 \big]
	&\quad t\text{-dependent exponential},		\\
&\hspace*{-0.35cm} &
 \hspace*{-0.05cm}(\Lambda^2-m_\pi^2)/(\Lambda^2-t)
	&\quad t\text{-dependent monopole},		\\
&\hspace*{-0.35cm} &
 \hspace*{-0.05cm}\big[ 1 - D_{\pi N}^2/(\Lambda^2-t)^2 \big]^{1/2}
	&\quad \text{Pauli-Villars (PV)},	           \\
&\hspace*{-0.35cm} &
 \hspace*{-0.05cm}\bar{x}_L^{-\alpha_\pi(t)}
	  \exp\big[ D_{\pi N}/\Lambda^2 \big]
	&\quad \text{Regge}.
\label{eq:models}
\end{array}
\right.
\end{equation}
For the $s$-dependent regulator,
\begin{equation}
s = \frac{k_T^2+M^2}{x_L} + \frac{k_T^2+m_\pi^2}{\bar{x}_L}
\end{equation}
corresponds to the invariant mass squared of the $\pi N$ system in the infinite momentum frame~\cite{Zoller:1991cb, Melnitchouk:1992yd, Holtmann:1996be}, while for the Regge regulator $\alpha_\pi(t) = \alpha'_\pi t$, with $\alpha'_\pi \approx 1$~GeV$^{-2}$. 
Each model involves a cutoff mass parameter $\Lambda$, which is fitted simultaneously with the pion PDF shape parameters. 
In previous JAM analyses~\cite{McKenney:2015xis, Barry:2018ort, Barry:2021osv}, the $s$-dependent exponential form was found to give the best description of the LN data; here we use this for the baseline model, but examine fits also with the other models in Eq.~(\ref{eq:models}).

In principle the semi-inclusive structure function $F_2^{\rm LN (4)}$ is a function of the 4 variables $x_B$, $Q^2$, $\bar{x}_L$, and $k_T^2$.
In practice, however, both the H1 and ZEUS data are integrated over specific ranges of the transverse momentum squared $k_T^2$, from $k_{T\rm min}^2$, typically taken to be zero, up to the maximum allowed in the detector setup, $k_{T\rm max}^2$.
One therefore defines the $k_T^2$-integrated LN structure function as~\cite{ZEUS:2002gig}
\begin{equation}
F_2^{\rm LN} \equiv F_2^{\rm LN (3)}(x_B,Q^2,\bar{x}_L) 
= \int_{k_{T\rm min}^2}^{k_{T\rm max}^2} \dd k_T^2\, 
F_2^{\rm LN (4)}\, (x_B,Q^2,\bar{x}_L,k_T^2).
\end{equation}
Similarly, we define the $k_T^2$-integrated $\pi N$ splitting function as
\begin{equation}
f_{\pi N}(\bar x_L) 
\equiv \int_{k_{T\rm min}^2}^{k_{T\rm max}^2} \dd k_T^2\, 
\tilde{f}_{\pi N}(\bar x_L,k_T^2).
\end{equation}

In addition to the LN structure function measurements by H1~\cite{H1:2010hym}, the ZEUS Collaboration also measured the ratio of LN to inclusive cross sections~\cite{ZEUS:2002gig},
\begin{equation}
r(x_B,Q^2,x_L)
= \frac{\dd^3\sigma^{\rm LN}/\dd x_B\, \dd Q^2\, \dd x_L}
     {\dd^2\sigma^{\rm inc}/\dd x_B\, \dd Q^2}
\,\Delta x_L,
\end{equation}
where $\Delta x_L$ is the bin size in $x_L$.
Here, the inclusive cross section $\dd^2\sigma^{\rm inc} / \dd x_B\, \dd Q^2$ is expressed in terms of the inclusive proton structure function, $F_2^p$, while the LN cross section $\dd^3\sigma^{\rm LN} / \dd x_B\, \dd Q^2\, \dd x_L \propto F_2^{\rm LN}$.

At large values of $x_L \sim 1$, pion exchange is expected to be the dominant contribution. 
Other processes, such as absorption effects and the exchange of other, heavier mesons, play increasingly important roles at smaller $x_L$. 
Instead of choosing a specific minimum value of $x_L$ above which one pion exchange is assumed, we systematically vary the minimum value of $x_L$ across different fits to investigate the kinematic region for which the data can be described within this framework.

\subsection{Data selection}
\label{sec:data}

Following the previous JAM pion studies~\cite{Barry:2018ort, Barry:2021osv}, we base the current analysis on the combination of pion--nucleus DY from the E615 and NA10 experiments and the LN data from H1 and ZEUS.
The E615 measurement comprises 61 points for a 252~GeV $\pi^-$ beam on a tungsten target, while the two NA10 beam energies, 194 and 286~GeV, contribute 36 and 20 points, respectively.
The data are presented as $\dd^2\sigma / \dd\sqrt{\tau}\, \dd x_F$, which is related to the cross section in Eq.~(\ref{eq:dyfactorization}) through a simple change of variables.
To avoid the $J/\psi$ and $\Upsilon$ resonance regions and the edges of phase
space, we restrict the dilepton invariant mass and Feynman variable to
$4.16 < Q < 7.68$~GeV and $0 < x_F < 0.9$~\cite{Conway:1989fs, NA10:1985ibr}.

For the LN electroproduction measurements at HERA, we include the semi-inclusive structure function $F_2^{\rm LN}$ data from H1, and the ratios of the LN to inclusive DIS cross sections from ZEUS.
The nominal JAM selection restricts data from both experiments to the forward region $\bar x_L < 0.2$, where the one-pion exchange mechanism is able to describe the data, yielding 58 H1 and 50 ZEUS points~\cite{H1:2010hym, ZEUS:2002gig}.
In this region the LN data extend the pion-PDF sensitivity to $x$ values significantly below those accessible by the DY data, and provide the principal information on the sea and gluon distributions.
We refer to this as the ``JAM baseline'' scenario.

The Fant\^{o}mas analysis~\cite{Kotz:2023pbu}, on the other hand, used, in addition to pion--nucleus DY and WA70 prompt-photon data~\cite{WA70:1987bai}, 29 points from H1 in the interval $0.23 < \bar x_L < 0.32$.
These points were fitted as an extracted pion structure function $F_2^\pi$ using a fixed integrated pion flux, with a 30\% uncorrelated uncertainty assigned to represent the model dependence.
By contrast, the JAM analyses fit the H1 and ZEUS LN observables by determining the pion--nucleon splitting function simultaneously with the pion PDFs.
To isolate the impact of the different treatments of the LN data, our Fant\^{o}mas-matched scenarios include the JAM DY dataset and theory calculation, but replace the nominal LN selection by the 29 H1 points and the associated 30\% uncertainty, which we refer to as the ``Fant\^omas H1 data'' selection.

\begin{table}[b]
\centering
\caption{Fit scenarios used for the LN data in the controlled comparisons, indicating the ranges of $\bar{x}_L$ covered for H1 and ZEUS and the corresponding number of points $N_{\rm pts}$ included. For the Fant\^omas H1 data selection, a 30\% systematic uncertainty was added to the data points~\cite{Kotz:2023pbu}. \\}
\label{tab:scenarios}
\begin{tabular}{@{}llc@{}}
\hline \hline
~Analysis       & ~~LN data selection         & ~~$N_{\rm pts}$~~ \\
\hline
~JAM baseline   & ~~H1\ $\bar{x}_L<0.2$       & 58 \\
                & ~~ZEUS\ $\bar{x}_L<0.2$     & 50 \\ \hline
~Fant\^{o}mas   & ~~H1\ $0.23<\bar{x}_L<0.32$ & 29 \\
                & ~~no ZEUS                   &  0 \\ \hline
~JAM extended   & ~~H1\ $\bar{x}_L<0.32$      & 87 \\
                & ~~ZEUS\ $\bar{x}_L<0.2$     & 50 \\ \hline\hline
\end{tabular}
\end{table}

Finally, we also consider a ``JAM extended scenario'' which retains the nominal ZEUS cut but combines the JAM baseline and Fant\^omas H1 data coverage to include the entire region $\bar x_L < 0.32$, for a total of 87 H1 and 50 ZEUS points.
This scenario tests whether the conclusions obtained with the JAM baseline remain stable when additional H1 information is included.
We summarize each of these scenarios in Table~\ref{tab:scenarios}.

\begin{figure}[!t]
\centering
\includegraphics[width=0.8\linewidth]{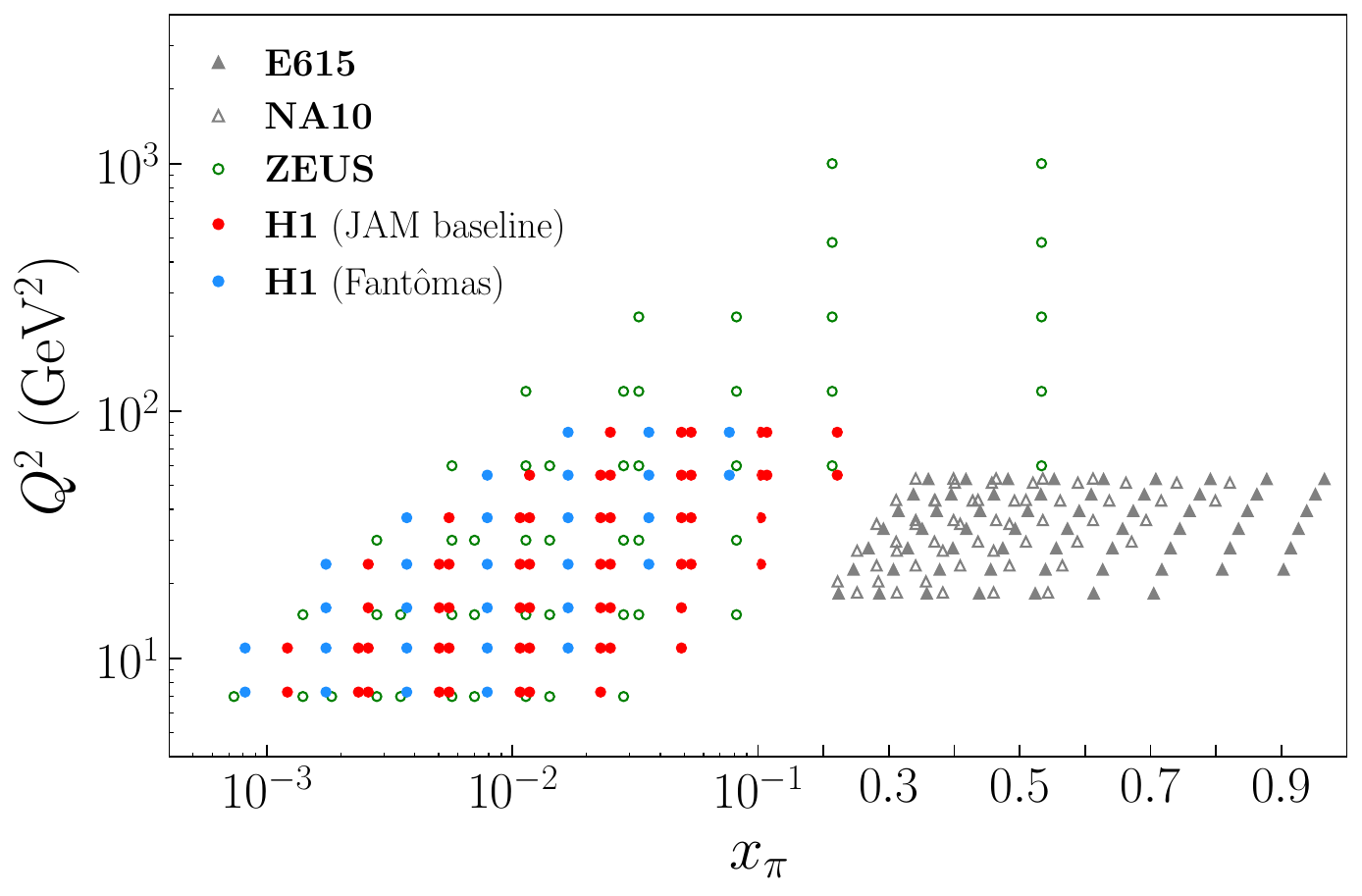}
\vspace*{-0.3cm}
\caption{Kinematic coverage in $(x_\pi,Q^2)$ of the pion-nucleus DY data from E615~\cite{Conway:1989fs} (filled triangles) and NA10~\cite{NA10:1985ibr} (open triangles) and the LN electroproduction data from H1~\cite{H1:2010hym} (solid circles) and ZEUS~\cite{ZEUS:2002gig} (open circles) used in the analysis. The H1 data are shown for the JAM baseline analysis (red circles) and Fant\^omas selection (blue circles).}
\label{fig:kinematics}
\end{figure}

The kinematical coverage of the DY and LN measurements is shown in Fig.~\ref{fig:kinematics}, illustrating the complementary roles of the DY and LN data.
The E615 and NA10 data predominantly cover moderate to large $\xpi$, where the pion valence distribution is largest, over a comparatively narrow range of $Q^2$.
The H1 and ZEUS measurements, in contrast, extend the analysis to substantially smaller $\xpi$ and span a broader $Q^2$ lever arm. 
This low-$\xpi$ coverage is essential for separating the sea and gluon contributions through QCD evolution.
It also explains why restricting the LN input to only 29 H1 points weakens the sea--gluon constraints, since the restricted sample retains only part of the kinematic information afforded by the combined H1 and ZEUS datasets, as we will discuss below.

\subsection{Parametrization}
\label{sec:param}

The pion PDFs are parametrized at the input scale $\mu=m_c=1.28~{\rm GeV}$ in terms of template functions defined by
\begin{equation}
T(x;\bm{\theta}) = x^a(1-x)^b\left(1+c\sqrt{x}+d x\right),
\label{eq:pdftemplate}
\end{equation}
where $\bm{\theta}=\{a,b,c,d\}$ denotes the set of shape parameters.
For all pion PDFs we impose isospin symmetry, 
    $u^{\pi^+}_v = \bar d^{\,\pi^+}_v = d^{\,\pi^-}_v = \bar u^{\pi^-}_v 
    \equiv q^\pi_v$,
and a flavor-symmetric light-quark sea, 
    $d^{\,\pi^+} = \bar u^{\pi^+} = s^{\pi^+} = \bar s^{\pi^+}
    \equiv q^\pi_s$.
The corresponding quark and antiquark PDFs are then given by
    $u^{\pi^+} = \bar d^{\,\pi^+} = q^\pi_v + q^\pi_s$.
The degrees of freedom to be reconstructed are then the valence quark, sea quark, and gluon distributions, which can be written as
\begin{subequations}
\label{eq:baselinepdf}
\begin{align}
 q_v^\pi(x,\mu) &= N_v\, T(x;\bm{\theta}_v),
 \\
 q_s^\pi(x,\mu) &= N_s\, T(x;\bm{\theta}_s),
 \\
 g^\pi(x,\mu)   &= N_g\, T(x;\bm{\theta}_g).
\label{eq:baselinegpdf}
\end{align}
\end{subequations}
The limited available data do not allow the polynomial shape parameters in Eq.~(\ref{eq:pdftemplate}) to be resolved in the sea and gluon sectors, and we therefore set $c_s = d_s = c_g = d_g = 0$.
The valence quark number and momentum sum rules fix two of the three normalizations ($N_v$~and~$N_s$).

To test the sensitivity to functional form flexibility, we also consider a two-shape parametrization for the valence and sea quark distributions,
\begin{equation}
q_i^\pi(x,\mu)
= N_i^{(1)}\, T(x;\bm{\theta}_i^{(1)})
+ N_i^{(2)}\, T(x;\bm{\theta}_i^{(2)}),
\qquad
\quad i=v,s,
\label{eq:extendedpdf}
\end{equation}
where $\bm{\theta}_i^{(2)}=\{a_i^{(2)},b_i^{(2)},0,0\}$, while retaining the one-shape form in Eq.~\eqref{eq:baselinegpdf} for the gluon.
We also perform fits both with and without positivity imposed on the quark PDFs over the fitted $x_\pi$ region at the input scale.
Comparing the unconstrained two-shape fits with the positivity constrained two-shape and one-shape fits allows us to distinguish genuine data driven uncertainty from broadening of the uncertainty bands resulting from the appearance of unphysical replicas.
Finally, to explore the possibility of a zero gluon PDF, suggested in one of the solutions found by the Fant\^{o}mas collaboration~\cite{Kotz:2023pbu}, we additionally perform a fit where the gluon PDF is fixed to zero at the input scale, and generated perturbatively through $Q^2$ evolution at higher scales.

\subsection{Bayesian analysis}
\label{sec:Bayesian}

Following previous JAM analyses~\cite{Cao:2021aci, Barry:2022aix}, we use an ensemble approach to reconstruct approximate posterior distributions for the pion PDF and splitting-function parameters.
Replicas are generated by Gaussian fluctuations of the data within their point-to-point uncorrelated uncertainties, and the parameters are refitted for each replica.
The Gaussian likelihood is proportional to $\exp(-\frac12\chi^2)$, with
\begin{equation}
\chi^2 = \sum_e \left\{
\sum_i \left[
\frac{d_i^e-t_i^e(\bm{\theta})/n_e-\sum_k r_k^e\beta_{ki}^e}{\alpha_i^e}
\right]^2
+ \sum_k (r_k^e)^2
+ \left(\frac{n_e-1}{\delta n_e}\right)^2
\right\},
\label{eq:fit-chi2}
\end{equation}
where $e$ denotes datasets, $d_i^e$ and $t_i^e$ are the data and theory values, $\alpha_i^e$ are uncorrelated uncertainties, and $\beta_{ki}^e$ are the effective correlated-error coefficients, scaled to the theory for multiplicative sources.
The correlated shifts $r_k^e$ are optimized analytically, while the normalization factors $n_e$ are fitted with the quoted uncertainties $\delta n_e$; the last two terms penalize departures from the nominal systematic shifts and normalizations.
In the datasets analyzed in this work, the ZEUS experiment is the only one that provides correlated uncertainties, while the others report only uncorrelated and normalization uncertainties.
In particular, the $0 \leq \bar{x}_L \leq 0.12$ bin has a somewhat sizable, 25\% energy scale uncertainty~\cite{ZEUS:2002gig}, for which the correlated shifts implemented in Eq.~(\ref{eq:fit-chi2}) are necessary to describe the data.
The resulting ensemble determines the PDF central values and uncertainty bands, with the PDF and splitting function parameters varied simultaneously.

For the post-fitting analysis, we assess the goodness of fit using the $\chi^2_{\rm red}$ and $Z$-score metrics.
The former is the Pearson $\chi^2/N_{\rm pts}$, where we compute the expectation value of the prediction (theory modified by the multiplicative normalization and added correlated shifts) to evaluate the first term in Eq.~(\ref{eq:fit-chi2}).
The $Z$-score is defined as $Z=\Phi^{-1}(p) \equiv \sqrt{2}\,{\rm erf}^{-1}(2p-1)$, where the $p$ value is computed with the Pearson $\chi^2$ relative to the expected $\chi^2$ distribution.

\section{Benchmarking with the Fant\^omas analysis}
\label{sec:results}

In this section, we benchmark our analysis against the Fant\^omas results by performing controlled fits that isolate the effects of LN data coverage, PDF parametrization flexibility, and positivity constraints.
We first examine the origin of the enlarged valence-PDF uncertainty found in  Ref.~\cite{Kotz:2023pbu}, and then assess the extent to which the restricted H1 dataset can constrain the pion gluon distribution.

\subsection{Origin of the enlarged valence PDF uncertainty}

\begin{figure}[t]
\centering
\includegraphics[width=0.92\linewidth]{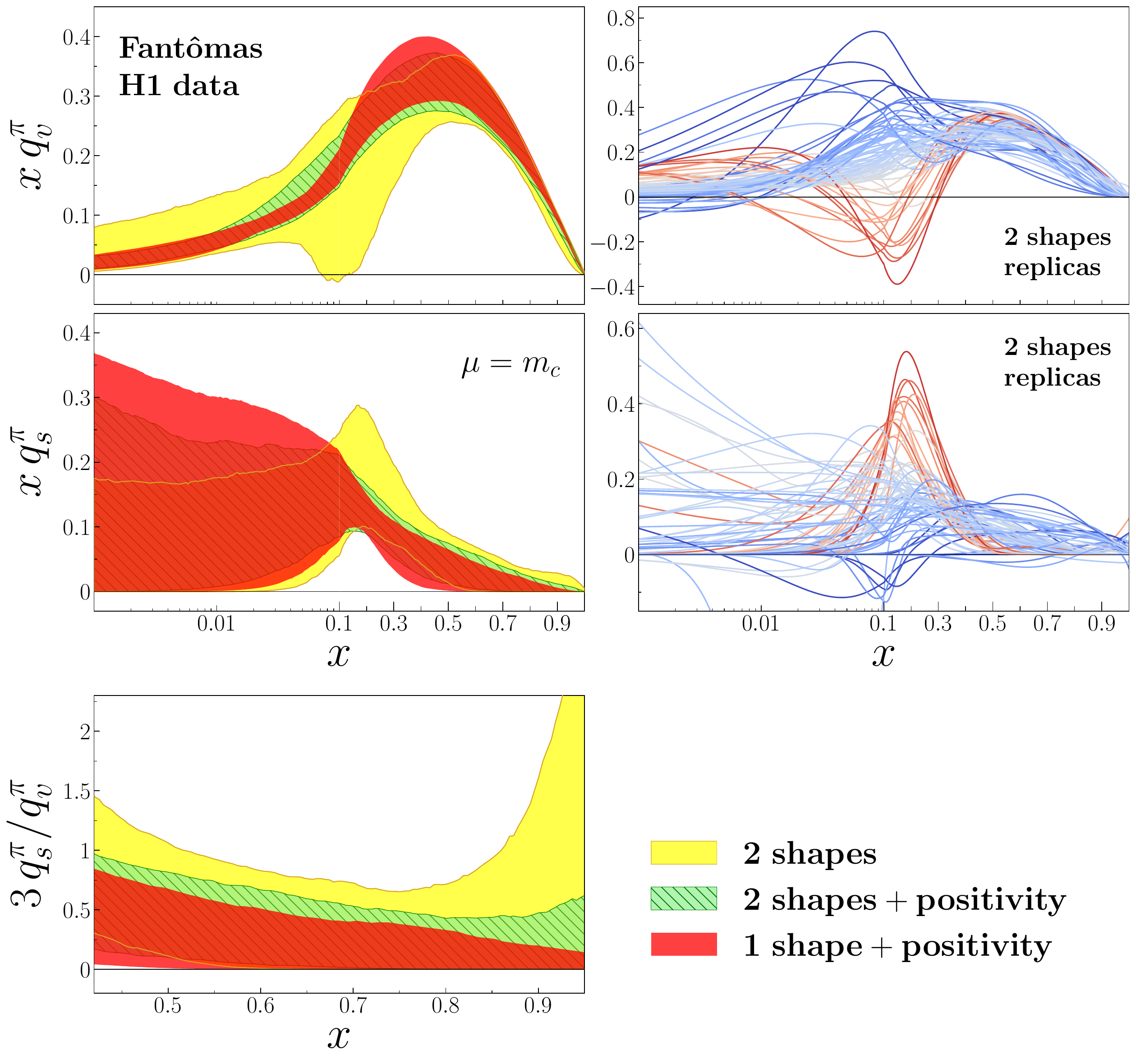}
\caption{Valence quark PDF (top row), sea quark PDF (middle row), and the sea to valence ratio (bottom row) at the input scale $\mu=m_c$ from fits to the data used in the Fant\^omas analysis~\cite{Kotz:2023pbu} (DY and restricted H1 dataset with an additional 30\% uncertainty), for the unconstrained 2 shapes parametrization (yellow 68\% CI bands), 2 shapes parametrization with positivity (green hatched bands), and 1 shape parametrization with positivity (red bands). The right columns show the individual replicas for the unconstrained 2 shapes parametrization, with the blue and red colored replicas indicating the correlations between the negative valence quark and strongly enhanced sea quark PDFs at $x \approx 0.1-0.2$. The factor of 3 in the sea to valence ratio accounts for the two valence quark versus six light sea-quark contributions in the pion.}
\label{fig:pdfs-two-shapes-Fantomas}
\end{figure}

As discussed in Sec.~\ref{sec:data} and summarized in Table~\ref{tab:scenarios}, the Fant\^{o}mas pion PDF analysis used a restricted 29-point selection of the H1 dataset, introducing an additional 30\% uncertainty on the pion flux, and no data from the ZEUS experiment.
We performed a fit to the H1 Fant\^{o}mas dataset, together with the standard DY data from the E615 and NA10 experiments, using several different parametrizations, including an unconstrained 2-shape parametrization, a 2-shape form with positivity constraints, and a 1-shape parametrization with positivity.
The resulting fits are illustrated in Fig.~\ref{fig:pdfs-two-shapes-Fantomas}.
All three scenarios yielded comparable global fit quality, with the 2 shapes without positivity and with positivity giving $\chi_{\rm red}^2$~($Z$-score) = $0.49~(-5.38)$ and $0.52~(-5.02)$, respectively, and the 1 shape with positivity having $0.53~(-4.81)$.
The figure shows that the unconstrained 2-shape parametrization (yellow bands) qualitatively reproduces the enlarged valence uncertainty around $x \approx 0.1$ that was reported in Ref.~\cite{Kotz:2023pbu}.
Further inspection of the individual replicas reveals that this broad band is driven by solutions for which the valence distribution becomes negative. 
Interestingly, the same replicas are correlated with an enhanced sea quark distribution at $x \approx 0.1-0.2$ relative to the valence quark PDF, and which significantly exceeds the valence at larger $x$ values.
A larger sea quark distribution compared with the valence in the large-$x$ region would be contrary to the orthodox physical picture of the pion being dominated by valence quarks at large $x$. 
The combination therefore of a negative valence PDF and a sea-dominated pion at large $x$ indicates that the additional parametrization freedom is exploring poorly constrained and physically disfavored solutions, rather than revealing a robust increase in the valence uncertainty.

These solutions can be suppressed in two independent ways. 
Firstly, imposing positivity on the 2-shape parametrization (green hatched bands) removes the negative replicas and substantially reduces the valence quark PDF uncertainty.
Alternatively, reducing the parametrization freedom to the positive 1-shape form (red bands) leads to nearly the same valence distribution and uncertainty as the 2-shape fit with the positivity constraint.
The agreement between these two constrained results suggests that the large uncertainty on the unconstrained fit is not a stable consequence of fits with the restricted LN datasets.

\begin{figure}[!t]
\centering
\includegraphics[width=0.92\linewidth]{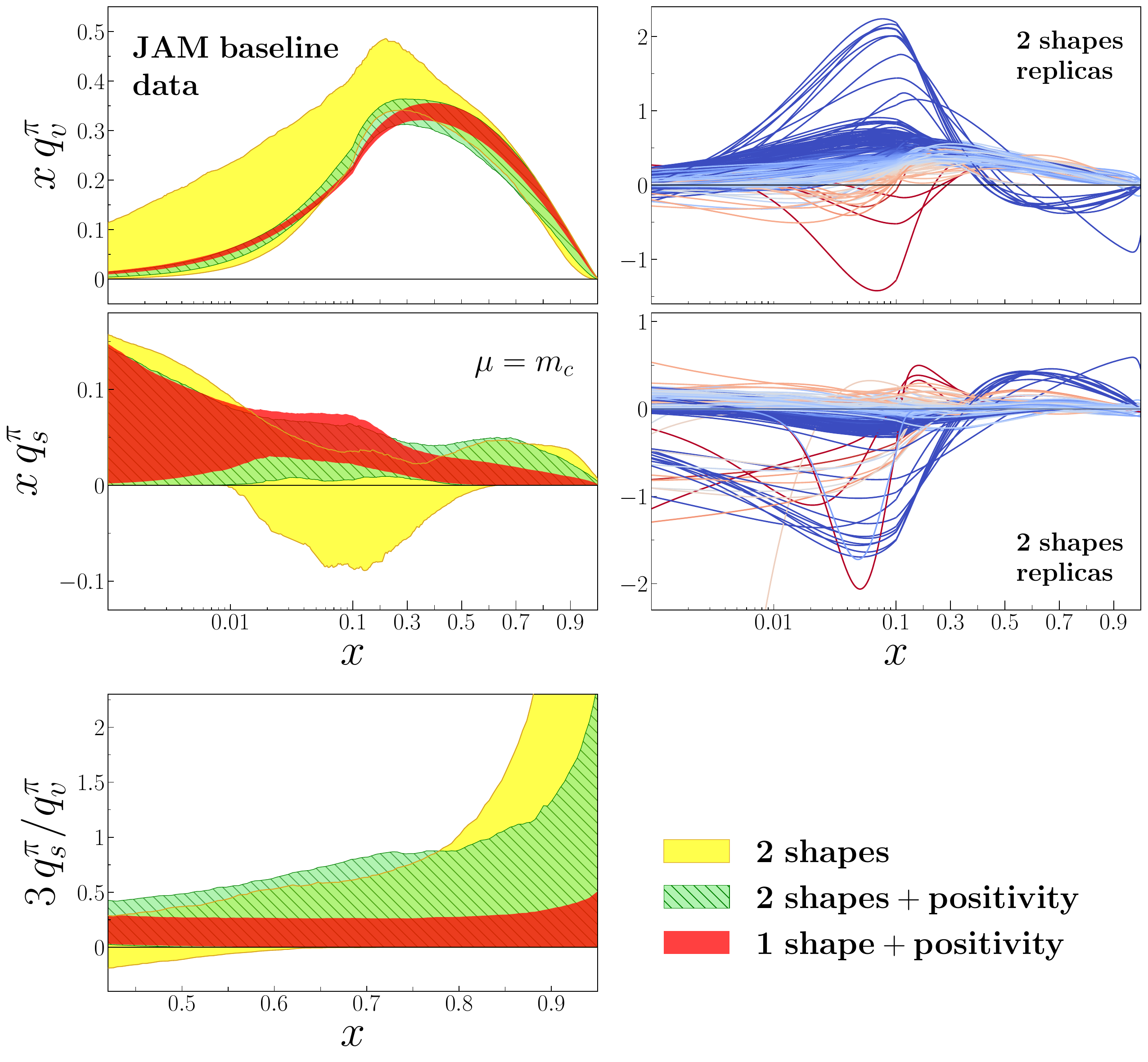}
\caption{As in Fig.~\ref{fig:pdfs-two-shapes-Fantomas}, but for the JAM baseline~\cite{Barry:2021osv} data selection. The enlarged small-$x$ valence   uncertainty in the unconstrained 2 shapes fit (yellow bands) is correlated with strongly negative sea-quark replicas (blue lines), which are removed by the positivity constraint.}
\label{fig:pdfs-two-shapes-JAM}
\end{figure}

Repeating the analysis using the baseline JAM data selection, which includes both H1 and ZEUS datasets, we illustrate the results in Fig.~\ref{fig:pdfs-two-shapes-JAM}.
Again, using 2 shapes unconstrained and 2 shapes with positivity gives comparable $\chi_{\rm red}^2$~($Z$-score) values of $0.83~(-1.90)$ and $0.78~(-2.48)$, respectively, while using a single shape with positivity gives $0.81~(-2.07)$.
The unconstrained 2-shape fit again generates an enlarged valence quark uncertainty (yellow bands) below $x \approx 0.1$.
In this case the broad valence solutions are strongly correlated with sea-quark PDF replicas that become large and negative.
Although negative PDFs are in principle allowed beyond leading order in $\alpha_s$, as long as observables remain positive, a large negative PDF will very likely lead to unphysical cross sections in some regions of phase space.
Enforcing positivity eliminates these replicas and sharply reduces the allowed variation (green hatched bands).
However, the ratio of sea-quark to valence-quark PDFs in this scenario remains large in the large-$x$ region, leading to the same interpretability problems of the fits in Fig.~\ref{fig:pdfs-two-shapes-Fantomas}.
Restricting the fit to the positive 1-shape form (red bands) produces a comparable result for the valence quark PDF, but without the anomalous sea-quark enhancement at high $x$.
For the remainder of this analysis we will therefore use the positive 1-shape parametrization as the simplest form that captures the data-supported valence-quark PDF uncertainty. \\ \\

\subsection{Gluon PDF constraints from LN data}

\begin{figure}[t]
\centering
\includegraphics[width=0.77\textwidth]{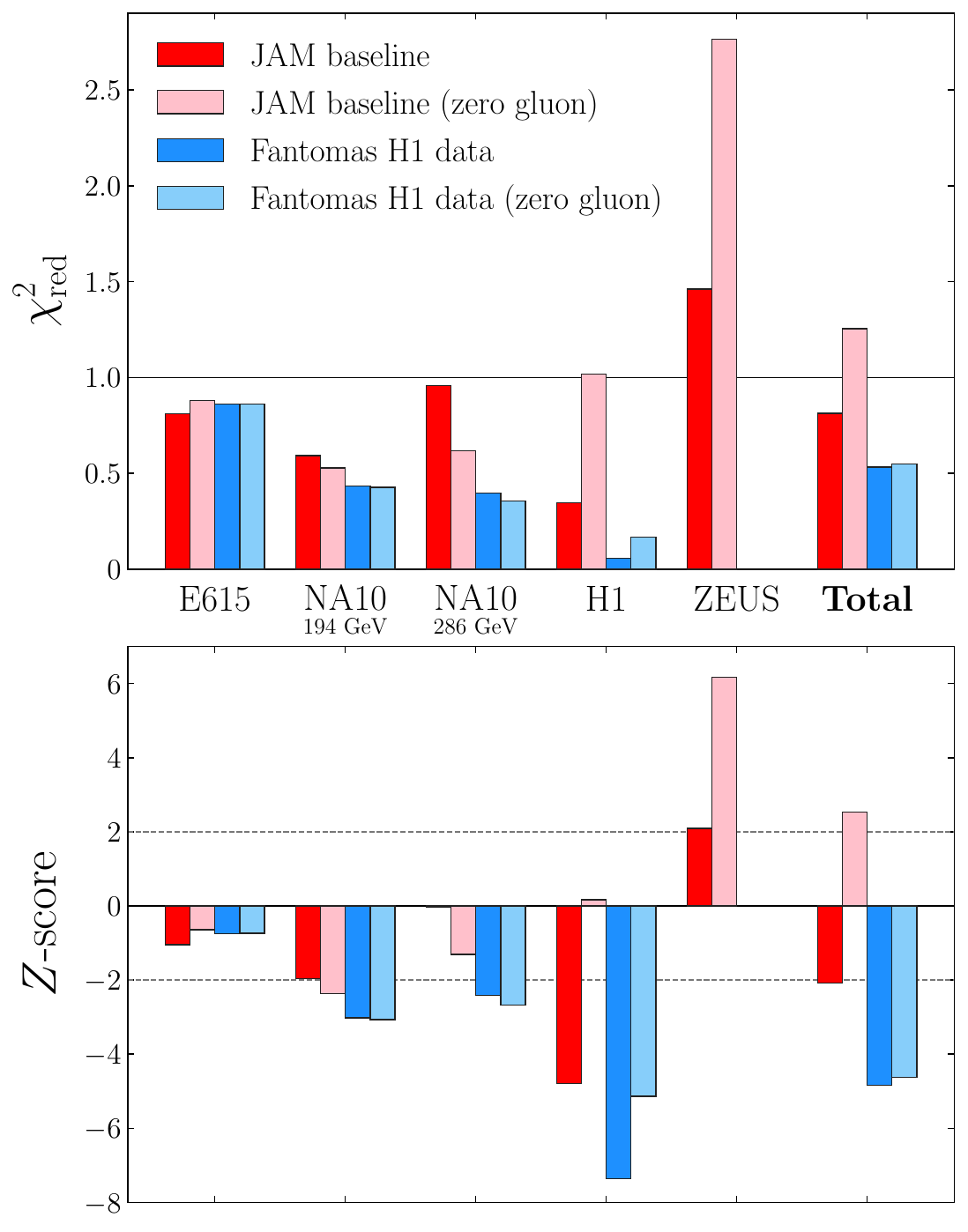}
\caption{Histograms of the $\chi^2_{\rm red}$ and $Z$-score values for each of the DY and LN datasets used in the global analysis, comparing the results of the JAM baseline fit (red) with a fit with zero gluon PDF at the input scale (pink), and with the corresponding fit using the Fant\^omas H1 data selection with a free gluon (blue) and with the gluon PDF set to zero at the input scale (light blue). Horizontal lines for $\chi^2_{\rm red}=1$ and $Z$-score $= \pm 2$ are shown as roughly optimal accepted metrics.}
\label{fig:zerogluon-fit-quality}
\end{figure}

\begin{figure}[t]
\centering
\includegraphics[width=0.75\linewidth]{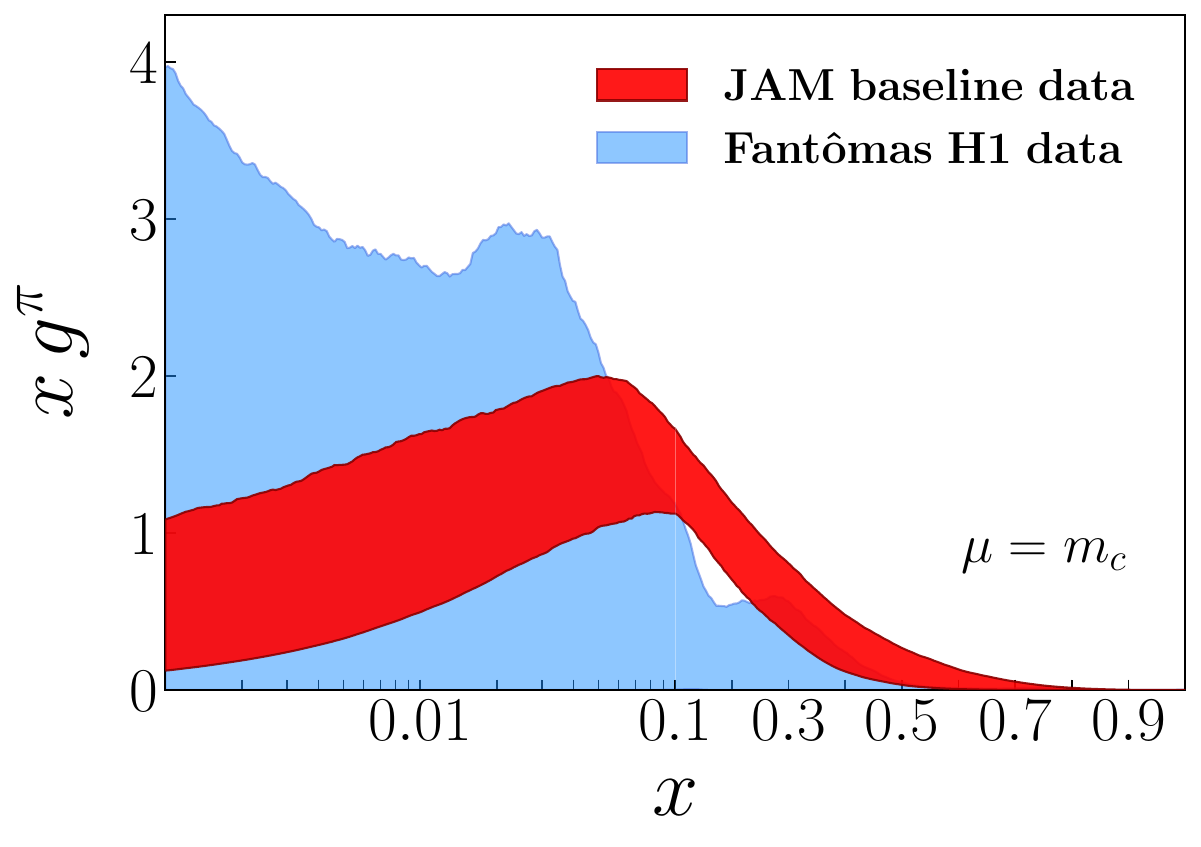}
\vspace*{-0.3cm}
\caption{Gluon PDF in the pion at the input scale $\mu=m_c$ fitted to the JAM baseline data (red 68\% CI band) and to the more restricted Fant\^omas H1 data selection (blue band) --- see Table~\ref{tab:scenarios}.}
\label{fig:zero-gluon-pdfs}
\end{figure}

To test the sensitivity of the gluon PDF in the pion to the LN data, we set the gluon distribution to zero at the input scale, $g^\pi(x,\mu)=0$, and refit all remaining PDF and splitting function parameters.
The results of the fits for the reduced $\chi^2_{\rm red}$ and $Z$-score values are summarized in Fig.~\ref{fig:zerogluon-fit-quality}.
We find that with the Fant\^{o}mas-matched data selection (blue/light blue histograms), the 29 H1 points and their additional 30\% uncertainty can accommodate a vanishing input gluon. 
In this case, fixing the input gluon to zero only changes the total $\chi^2_{\rm red}$ from 0.53 to 0.55, corresponding to $\Delta\chi^2 \approx 2$ for the 146 fitted points. 
The restricted dataset clearly has little power to distinguish the free-gluon and zero-gluon hypotheses, and we are able to reproduce, within the JAM framework, the essential Fant\^{o}mas result that a zero-gluon solution is allowed by this restricted LN input.
The $x$ dependence of the fitted gluon PDF in the pion, $x g^\pi$, is illustrated in Fig.~\ref{fig:zero-gluon-pdfs} at the input scale, $\mu=m_c$.

For the baseline JAM data selection (red/pink histograms in Fig.~\ref{fig:zerogluon-fit-quality}), on the other hand, the total $\chi^2_{\rm red}$ rises from 0.81 to 1.26, or by approximately $\Delta\chi^2=99$ for 225 points.
The deterioration is clearly concentrated in the LN measurements. 
Specifically, the H1 contribution to $\chi^2_{\rm red}$ increases from 0.35 to 1.02 per datum, while the ZEUS contribution increases from 1.46 to 2.76 per datum, and reaches a $Z$-score greater than 6.
The large ZEUS penalty, in particular, provides independent small-$x_\pi$ and large-$Q^2$ constraints that are absent in the Fant\^{o}mas analysis.
The combined H1 and ZEUS data, therefore, strongly disfavor a vanishing gluon at the input scale, and give rise to the nonzero shape for the JAM baseline data scenario illustrated in Fig.~\ref{fig:zero-gluon-pdfs}.

\begin{figure}[!]
\centering
\includegraphics[width=0.92\linewidth]{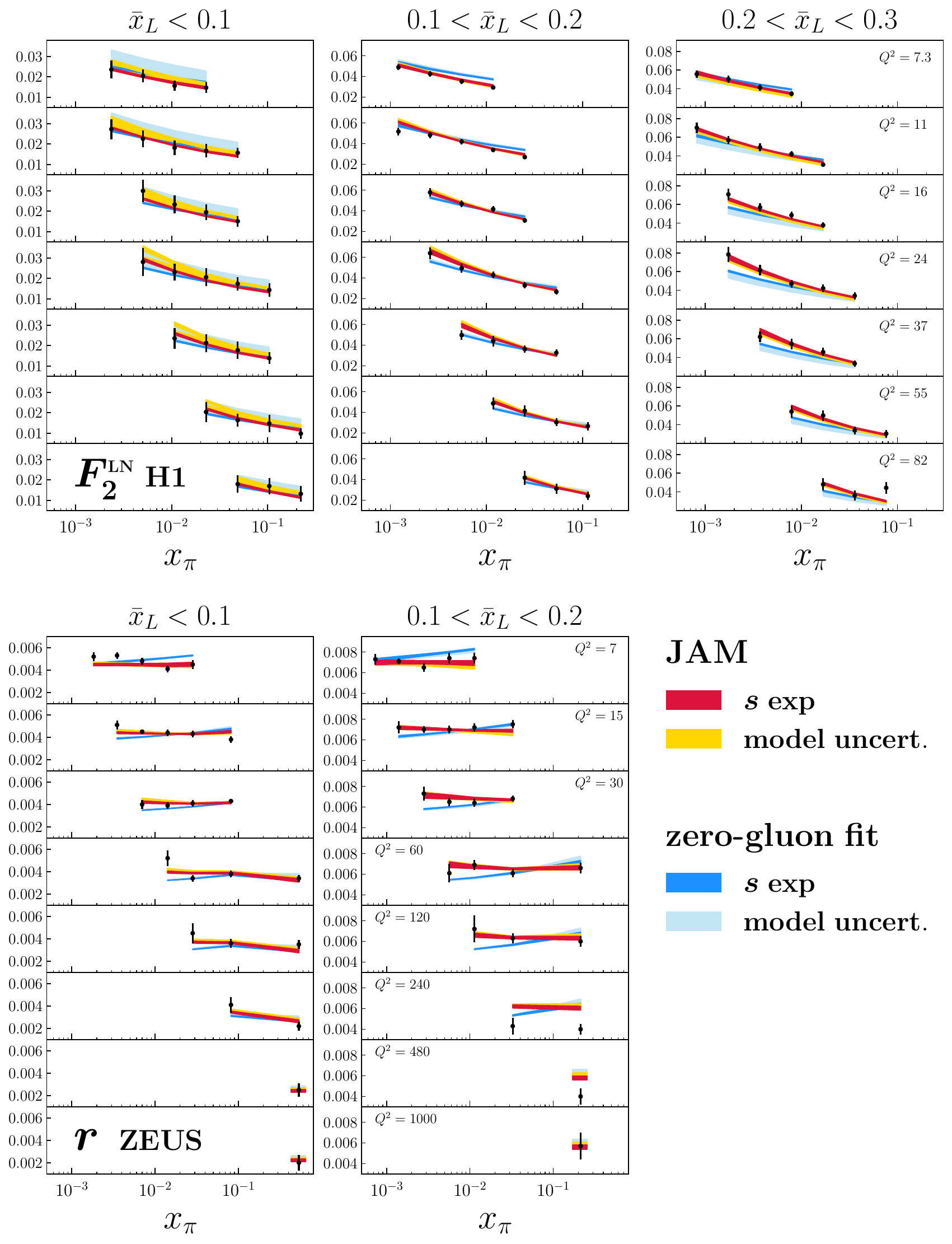}
\vspace*{-0.2cm}
\caption{LN structure function data from H1 ($F_2^{\rm LN}$) and ZEUS ($r$) versus $x_\pi$ for various $\bar x_L$ and $Q^2$ (in GeV$^2$) compared with the JAM results (red bands) and with fits with zero gluon PDF at the input scale (blue bands), using the favored $s$-dependent exponential splitting function model, with the envelopes from fits with other models indicated (yellow and light blue bands, respectively).}
\label{fig:data-vs-theory-H1ext}
\end{figure}

The origin of the $\chi^2_{\rm red}$ deterioration is clearly visible in Fig.~\ref{fig:data-vs-theory-H1ext}, where we compare the observed $x_\pi$ dependence of the LN cross sections across different H1 and ZEUS kinematic bins in $\bar x_L$ and $Q^2$ for the free-gluon and zero-gluon scenarios.
Overall, the free-gluon results (red bands) closely follow the observed $x_\pi$ dependence of the H1 and ZEUS data, whereas the zero-gluon results (blue bands) have slopes that are not as well aligned with the data trend.
The tension between the zero-gluon fits and the data is visible for both the H1 $F_2^{\rm LN}$ data, where for larger $\bar x_L$ the $x_\pi$ slope is shallower than observed, and the ZEUS $r$ ratio, where the $x_\pi$ slope is steeper than the data.
Note that the results shown in Fig.~\ref{fig:data-vs-theory-H1ext} are for the JAM extended analysis, which is discussed in detail in Sec.~\ref{sec:extensions}, as it includes the H1 data for $0.2 < \bar{x}_L < 0.3$; however, the arguments regarding the deterioration of the $\chi^2_{\rm red}$ remain.
As discussed in Sec.~\ref{sec:setup}, for the central results we use the $s$-dependent exponential regulator in Eq.~(\ref{eq:models}), while the uncertainty associated with using different regulators is illustrated in Fig.~\ref{fig:data-vs-theory-H1ext} for the free-gluon (yellow bands) and zero-gluon (light blue bands) results.
Except for the lowest $\bar x_L$ H1 dataset, the size of the regulator model uncertainty is smaller than the differences between the free-gluon and zero-gluon fit results.


\section{Extended JAM global QCD analysis}
\label{sec:extensions}

Having established the importance of broad LN coverage for constraining the pion gluon PDF, we now perform an extended JAM global QCD analysis that supplements the baseline JAM dataset with the H1 data at $0.23 < \bar{x}_L < 0.32$ that were the focus of the Fant\^{o}mas analysis.
We assess the robustness of the analysis by fitting with various models of the $\pi N$ splitting function in Eq.~(\ref{eq:models}) and examining the kinematic region in which the pion-exchange description provides a good description of the data.
The resulting splitting functions are further tested against independent $pp \to nX$ data at low $k_T$ and $\bar{x}_L$ values where one-pion exchange is expected to dominate.
Finally, we then present the pion PDFs using the extended dataset and quantify their sensitivity to the splitting function model.

\subsection{Splitting function systematics and kinematic coverage}
\label{sec:fluxchecks}

In the following we examine the ability of the JAM methodology to extend the analysis of the LN data to include the 29 H1 data points in the $0.2 < \bar{x}_L < 0.32$ region that were not fitted in the baseline JAM analysis.
To better understand and quantify the dependence of the analysis on the choice of splitting function, we vary the functional form of the $p \to \pi^+ n$ splitting function regulator and assess the quality of the fit for each model.
As suggested by the generally good agreement between data and theory in Fig.~\ref{fig:data-vs-theory-H1ext}, and indicated by the $\chi^2_{\rm red}$ and $Z$-scores in Table~\ref{tab:chi2_zscores_H1ext}, we find that H1 data are well-described by all models over the range of $\bar x_L$ considered.
However, similarly extending the ZEUS data results in larger $\chi^2_{\rm red}$ and $Z$-score values that indicate the higher-$\bar{x}_L$ ZEUS data cannot be described well in the one-pion exchange framework.
This may reflect the different ranges of $k_T$ over which the H1 and ZEUS data are integrated (see Fig.~\ref{fig:chiral_kinematics} below).
In the extended JAM analysis we therefore adopt the cuts $\bar{x}_L < 0.32$ for H1 and $\bar{x}_L < 0.2$ for the ZEUS LN data.

\begin{table}[b]
\centering
\footnotesize
\setlength{\tabcolsep}{4pt}
\caption{Values of $\chi^2_{\rm red}$ ($Z$-score) for each DY and LN dataset in the full JAM fit for the extended datasets, for each splitting function model defined in Eq.~(\ref{eq:models}), along with the number of data points $N_{\rm pts}$ for each dataset.\\}
\begin{tabular}{lcccccc}
\hline\hline
Experiment & ~~$N_{\rm pts}$~~ &
\multicolumn{5}{c}{$\chi^2_{\rm red}$ ($Z$-score)} \\
\cline{3-7}
& &
{\bfseries $\bm{s}$ exp}~ &
$t$ exp~ &
$t$ mon~ &
PV~ &
Regge \\
\hline
E615 & 61 &
$\bm{0.81~(-1.05)}$~ &
0.81 ($-1.07$)~ &
0.81 ($-1.06$)~ &
0.81 ($-1.07$)~ &
0.81 ($-1.08$) \\

NA10 (194 GeV) & 36 &
$\bm{0.60~(-1.94)}$~ &
0.64 ($-1.71$)~ &
0.63 ($-1.73$)~ &
0.60 ($-1.89$)~ &
0.63 ($-1.73$) \\

NA10 (286 GeV) & 20 &
$\bm{0.96~(-0.03)}$~ &
1.10 ($+0.41$)~ &
1.08 ($+0.35$)~ &
1.04 ($+0.22$)~ &
1.07 ($+0.33$) \\

H1 & 87 &
$\bm{0.41~(-5.04)}$~ &
0.77 ($-1.58$)~ &
0.81 ($-1.28$)~ &
0.82 ($-1.21$)~ &
0.85 ($-0.96$) \\

ZEUS & 50 &
$\bm{1.47~(+2.13)}$~ &
1.66 ($+2.83$)~ &
1.61 ($+2.67$)~ &
1.76 ($+3.19$)~ &
1.49 ($+2.20$) \\

\hline
Total & 254 &
$\bm{0.79~(-2.58)}$~ &
0.96 ($-0.40$)~ &
0.96 ($-0.37$)~ &
0.99 ($-0.08$)~ &
0.95 ($-0.50$) \\
\hline\hline
\end{tabular}
\label{tab:chi2_zscores_H1ext}
\end{table}

\begin{figure}[t]
\centering
\includegraphics[width=0.9\linewidth]{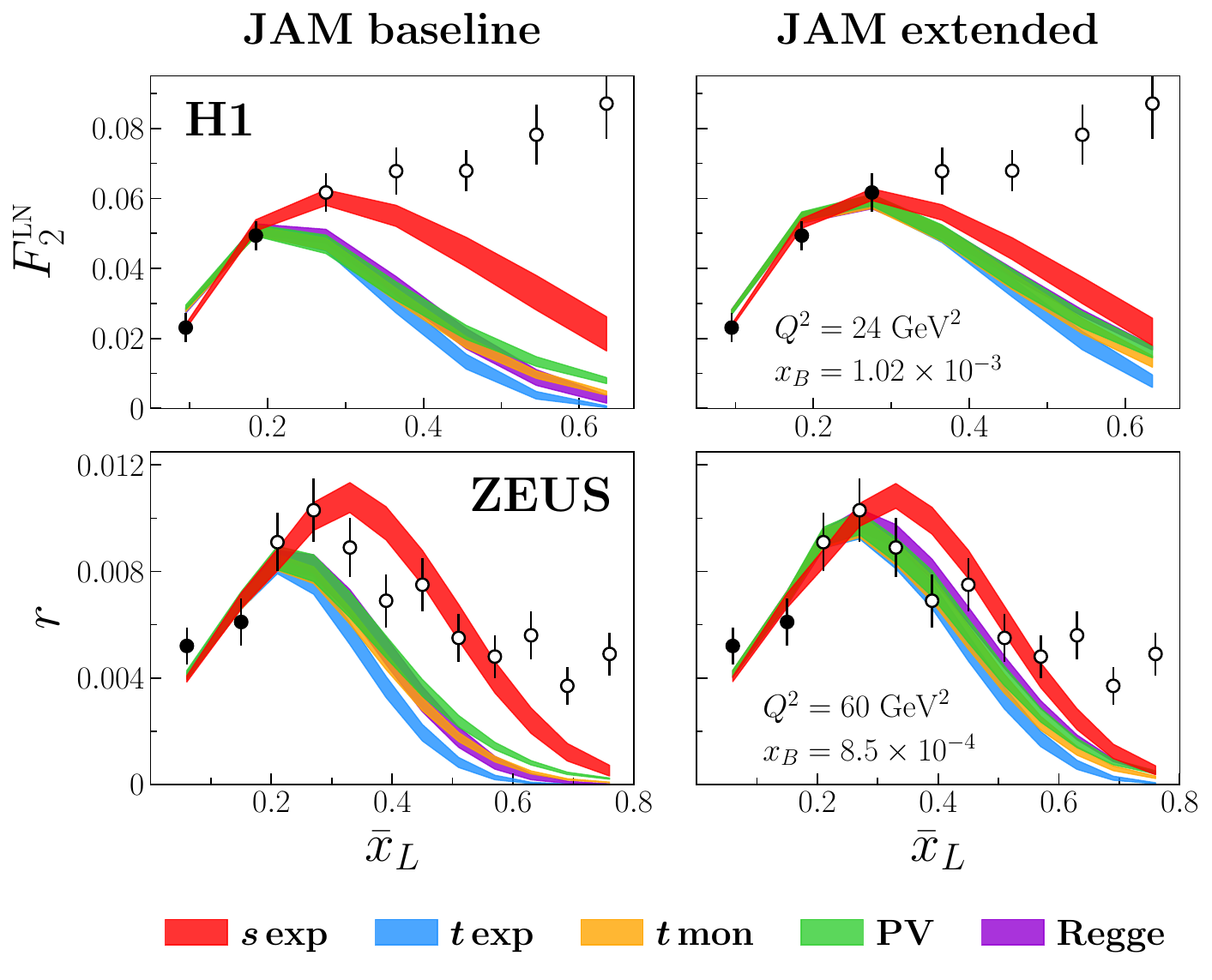}
\vspace*{-0.3cm}
\caption{Representative H1 (top) and ZEUS (bottom) LN data versus $\bar x_L$, for fixed $Q^2$ and $x_B$, compared with fitted JAM results using the baseline data ($\bar x_L < 0.2$ for H1 and ZEUS, left column) and extended data ($\bar x_L < 0.32$ for H1, right column), for each of the splitting function models in Eq.~(\ref{eq:models}). Filled (open) circles indicate data points that are (are not) included in the fits.}
\label{fig:data-vs-y}
\end{figure}

The comparison in Table~\ref{tab:chi2_zscores_H1ext} shows that the best fits are obtained for the $s$-dependent exponential regulator model, with the strongest preference indicated by the H1 data ($\chi^2_{\rm red} \approx 0.4$, compared with $\approx 0.8$ for the other models).
The ZEUS data are generally less well described, with larger $\chi^2_{\rm red} \approx 1.5-1.7$ values, although the $s$-dependent exponential regulator is still preferred, with $\chi^2_{\rm red} \approx 1.5$.
The fitted values for the regulator cutoff masses are listed in Table~\ref{tab:lambda_values}.
Interestingly, the cutoff mass for the $s$-dependent exponential model is relatively stable between the JAM baseline and JAM extended analyses, with $\Lambda \approx 1.6$~GeV.
In contrast, for each of the other models the cutoff masses required to fit the JAM extended dataset are 30\%--50\% larger than for the JAM baseline.
This can be understood from the shapes of the splitting functions and corresponding LN structure functions versus $\bar x_L$, as shown in Fig.~\ref{fig:data-vs-y} for representative datasets at $(x_B,Q^2) = (1.02 \times 10^{-3}, 24~{\rm GeV}^2)$ and $(8.5 \times 10^{-4}, 60~{\rm GeV}^2)$.
Since the results for the $t$-dependent forms (exponential, monopole, PV and Regge) fall faster with $\bar x_L$ than those with the $s$-dependent exponential form, to describe the $\bar x_L \approx 0.3$ H1 data points requires overall enhancement of the splitting function at higher $\bar x_L$, which can be achieved only through a larger cutoff mass.
In contrast, the splitting function for the $s$-dependent exponential regulator is broader at large $\bar x_L$ (although more strongly suppressed at lower $\bar x_L$), which enables the lower-$\bar x_L$ and higher-$\bar x_L$ points to be described for the same cutoff.
A similar effect is visible also for the ZEUS data, with the broader $s$-dependent exponential shape able to accommodate higher-$\bar x_L$ values, whereas a similar description of the higher-$\bar x_L$ data with the $t$-dependent forms would require a correspondingly larger cutoff mass.

\begin{table}[t]
\centering
\renewcommand{\arraystretch}{1.2}
\setlength{\tabcolsep}{8pt}
\caption{Fitted values of the cutoff parameters $\Lambda$ (in GeV) for each of the models in Eq.~(\ref{eq:models}) for the JAM baseline and JAM extended data selections.\\}
\begin{tabular}{lccccc}
\hline\hline
~Analysis & \multicolumn{5}{c}{$\Lambda$~(GeV)}
\\
\cline{2-6}
          &
{\bf $s$ exp} & $t$ exp & $t$ mon & PV & Regge~ 
\\
\hline
~JAM baseline &
{\bf 1.59(18)} & 0.64(4) & 0.57(4) & 0.26(2) & 0.92(12)
\\
~JAM extended &
{\bf 1.62(16)} & 0.98(6) & 0.91(6) & 0.42(3) & \!\!2.00(3) 
\\
\hline\hline
\end{tabular}
\label{tab:lambda_values}
\end{table}

\begin{figure}[t]
\centering
\includegraphics[width=\linewidth]{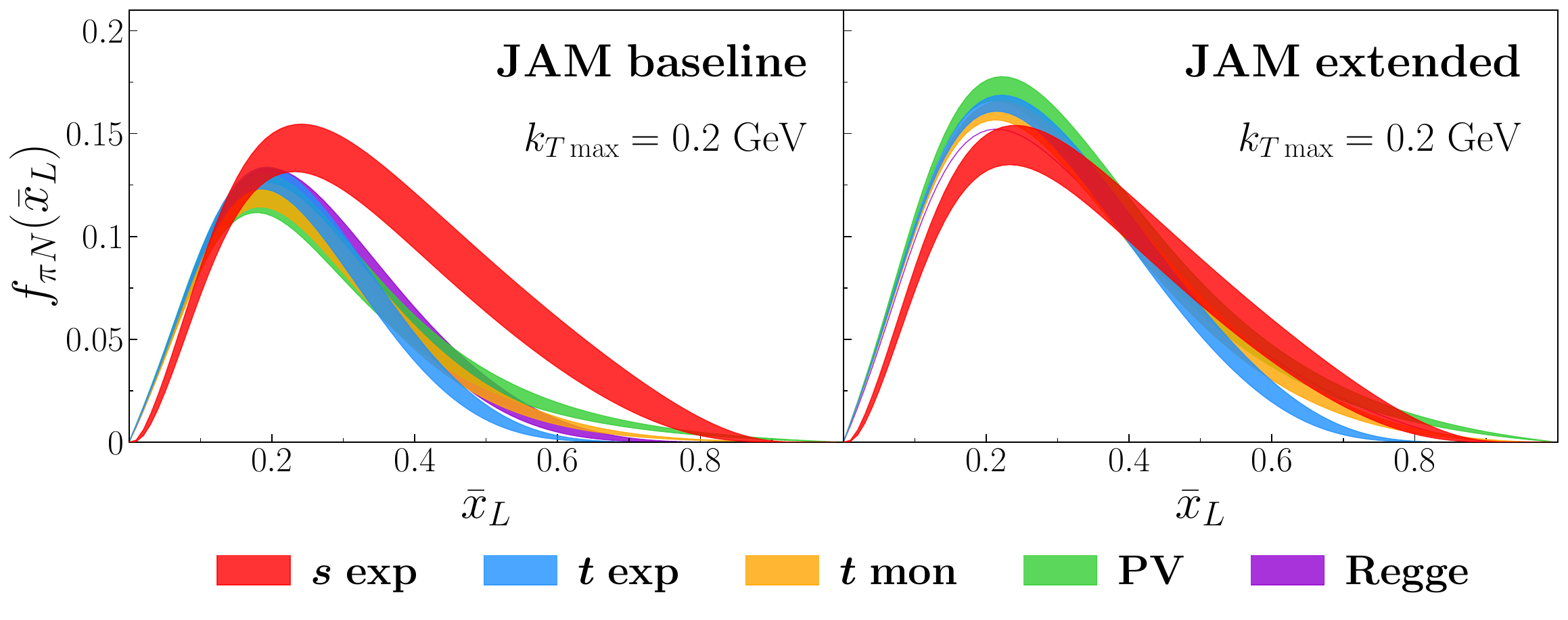}
\vspace*{-1cm}
\caption{Comparison of the fitted pion--nucleon splitting functions $f_{\pi N}(\bar x_L)$ for various regulator models at $k_{T\rm max}=0.2~\mathrm{GeV}$, obtained from fits to the JAM baseline (left) and JAM extended (right) data selections. \\}
\label{fig:splitting-functions}
\end{figure}

The corresponding fitted splitting functions, integrated up to $k_T=0.2$~GeV, are shown in Fig.~\ref{fig:splitting-functions} for each of the five models for the  baseline and extended scenarios.
The models give very similar predictions for the data in the fitted region, showing that the inferred kinematic limits are not an artifact of a particular regulator choice.
However, going from the JAM baseline data selection to the extended data selection, the $s$-dependent exponential model remains fairly stable, while the other models vary significantly depending on the H1 kinematic cut.

As noted above, the H1 and ZEUS data span over somewhat different regions of $k_T$ (or~$t$) --- for H1 the range is $k_T < 0.2$~GeV, while for ZEUS the upper limit on $k_T$ depends on $x_L$, $k_T < (0.656\, x_L)$~GeV.
For each $\bar x_L$ value considered in our analysis, the range of $t$ in Fig.~\ref{fig:chiral_kinematics} shows that the values of $-t$ are generally $< 0.1$~GeV$^2$ for H1, but can reach up to $\approx 0.45$~GeV$^2$ for the ZEUS data at the lowest $\bar x_L$.
At such values one expects one-pion exchange to be the dominant mechanism, especially for the H1 data, although this ultimately must be determined empirically.

\begin{figure}[t]
\centering
\includegraphics[width=1\linewidth]{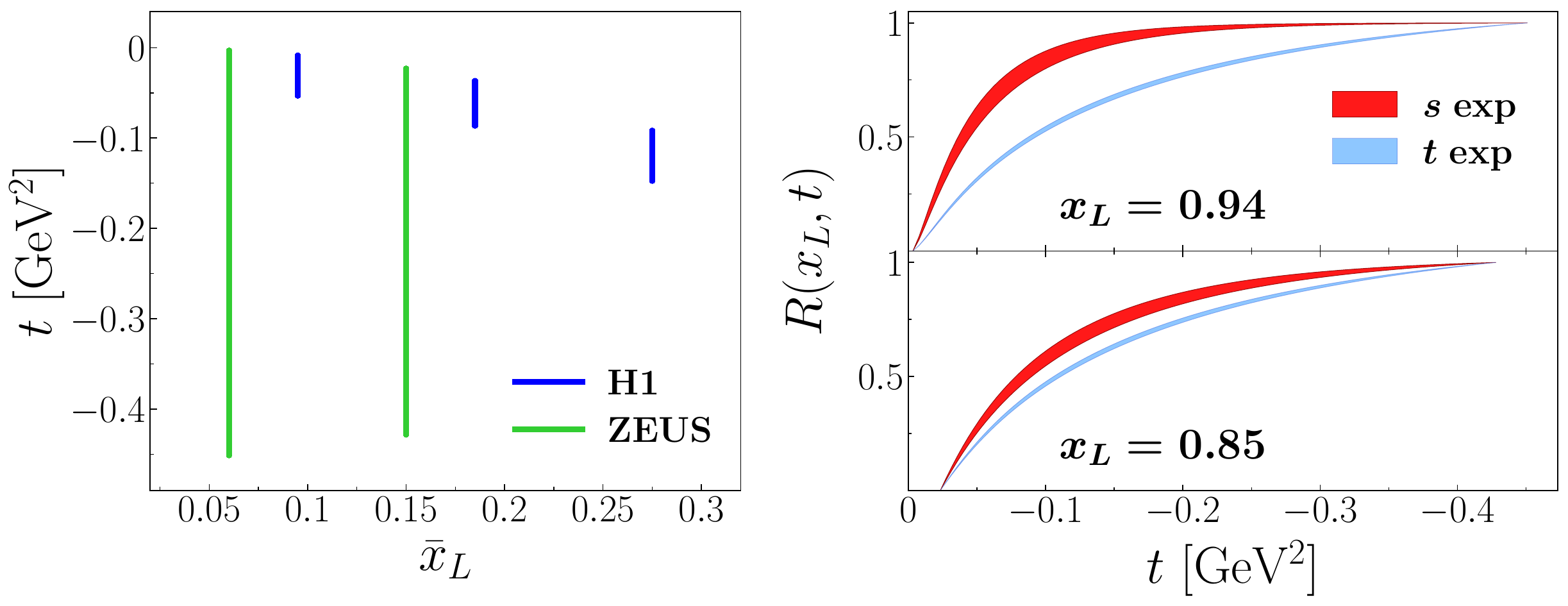}
\caption{
(Left) Kinematic ranges of the H1 (blue lines) and ZEUS (green lines) datasets of $t$ for fixed $\bar x_L$ values.
(Right) Fractional truncated integrals of the splitting functions, defined in Eq.~(\ref{eq.truncintegral}), as a function of $t$ for $x_L=0.94$ (top) and 
$x_L=0.85$ (bottom) at the ZEUS kinematics for the $s$-dependent exponential (red bands) and $t$-dependent exponential (blue bands) models.\\}
\label{fig:chiral_kinematics}
\end{figure}

A complementary way of assessing the relevance of the one-pion exchange mechanism at finite $k_T$ or $t$ is through the fractional truncated integral, defined by
\begin{equation}
R(x_L,t) 
= \frac{x_L}{f_{\pi N}(\bar{x}_L)}
\int_t^{t_0} \dd\hat{t}\ \tilde{f}_{\pi N}(\bar{x}_L, k_T^2(\hat{t})),
\label{eq.truncintegral}
\end{equation}
where $t_0 = -\bar{x}_L^2 M^2/x_L$ is the value of $t$ when $k_T^2=0$, and inverting Eq.~(\ref{eq:t}) one has \mbox{$k_T^2(t) = -x_L t - \bar{x}_L^2 M^2$}.
This ratio gives the cumulative contribution to the integrated splitting function, and is shown in Fig.~\ref{fig:chiral_kinematics} for $x_L=0.85$ and 0.94, for the $s$-dependent and $t$-dependent exponential regulators.
The denominator in Eq.~(\ref{eq.truncintegral}) is evaluated according to the ZEUS kinematic range, integrating from $k_{T\rm min}=0$ to the corresponding $k_{T\rm max}$ for each value of $\bar{x}_L$.
The result shows that while the ZEUS observable extends beyond the $t$ range for H1, the majority of the signal contributing to these observables is largely coming from the region where $t>-0.15~{\rm GeV}^2$, especially for the $s$-dependent regulator.

In a recent analysis, Chang, Hsieh, and Puhan~\cite{Chang:2026tgb} reported that, with the $s$-dependent exponential regulator, the one-pion exchange contribution can underestimate the ZEUS data at $x_L=0.94$ ($\bar{x}_L=0.06$) by as much as a factor of three. 
Their comparison, however, uses the published central values without the correlated systematic shifts and normalization factors that are fitted and penalized as nuisance parameters in the JAM analysis --- see Eq.~(\ref{eq:fit-chi2}).
The unshifted discrepancy therefore does not correspond to the contribution of this bin to the profiled JAM $\chi^2$, although it does indicate enhanced sensitivity to the pion-flux parametrization and the treatment of systematic uncertainties near $\bar{x}_L \approx 0$.

To quantify this sensitivity, we repeated the positive one-shape fits after removing the 25 ZEUS points at $\bar{x}_L=0.06$ and refitting the PDFs, splitting function, and nuisance parameters.
In the baseline free-gluon fit, the ZEUS data contribution decreases from $\chi^2_{\rm red}=1.42$ to 1.09, while the total correlated-systematic penalty decreases from 31.2 to 11.4.
The median cutoff for the $s$-dependent exponential regulator decreases from $\Lambda=1.55$~GeV to 1.28~GeV, accompanied by a decrease in the gluon momentum fraction from 0.42 to 0.30 and an increase in the sea-quark momentum fraction from 0.11 to 0.20.
Nevertheless, after removing this bin, setting the input gluon to zero still increases the total $\chi^2$ relative to the free-gluon fit by $\Delta\chi^2=85.4$ for the baseline dataset and 131.2 for the extended H1 dataset, including the correlated-shift and normalization penalties defined in Eq.~(\ref{eq:fit-chi2}).
The lowest-$\bar{x}_L$ ZEUS data therefore affect the quantitative sea--gluon separation, but not the preference for a nonzero input gluon.

\begin{figure}[t]
\centering
\includegraphics[width=0.96\linewidth]{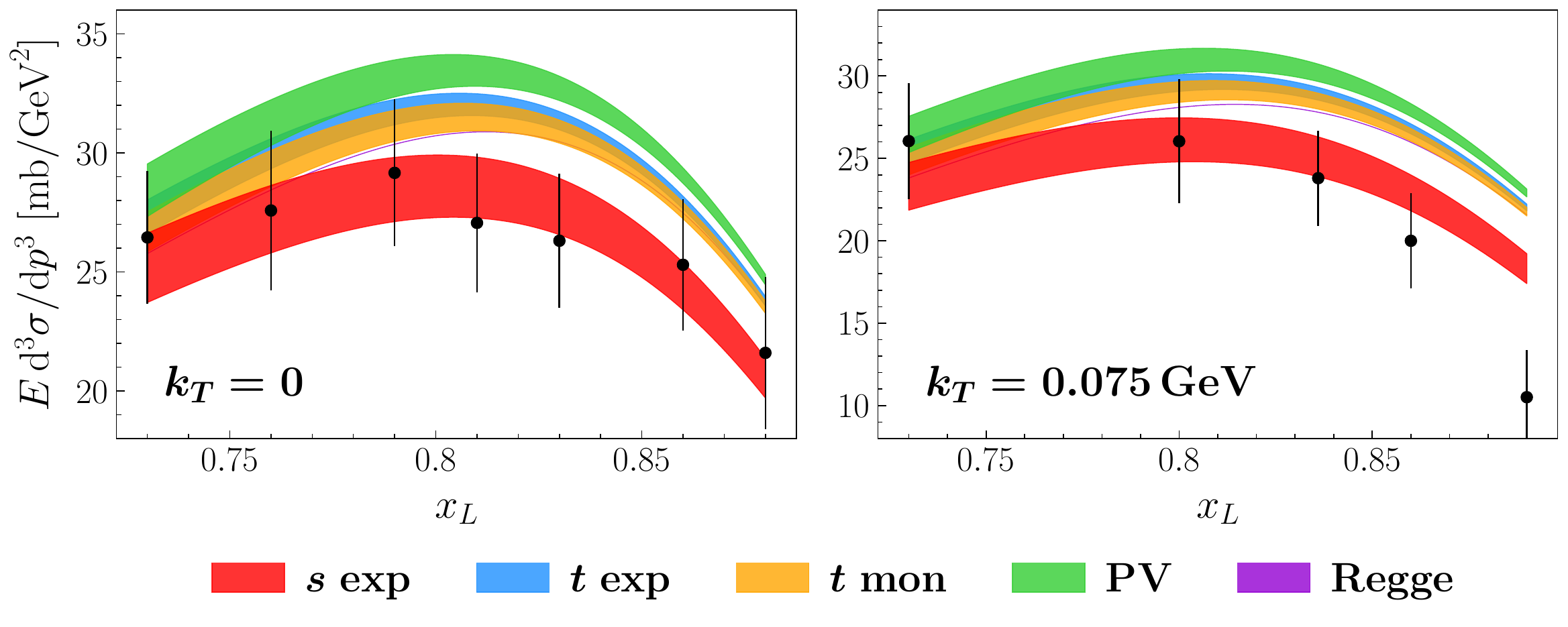}
\vspace*{-0.3cm}
\caption{Comparison of the $pp \to nX$ differential cross section versus $x_L$ for $k_T=0$ (left panel) and $k_T=0.075$~GeV (right panel) computed from the fitted splitting functions in the JAM extended data analysis for various regulators, at the 68\% credible interval. The data are from Refs.~\cite{Flauger:1976ju, Blobel:1978yj}.}
\label{fig:pp_to_nX}
\end{figure}

As an independent check of the physical content of the fitted $\pi N$ splitting function in our analysis, and the approximate universality of the pion exchange mechanism at low $k_T$, in Fig.~\ref{fig:pp_to_nX} we show the predictions for the inclusive $pp \to nX$ charge exchange cross sections from Refs.~\cite{Flauger:1976ju, Blobel:1978yj} using the fitted splitting functions from the JAM extended data analysis.
The calculated cross sections agree remarkably well, especially at the low values of $k_T$ where the one-pion exchange is expected to dominate.
The agreement is particularly striking for the $s$-dependent exponential regulator, for which the Pearson $\chi^2$ per datum value is 0.84, compared with the values \{2.86, 2.52, 4.05, 2.23\} for the \{$t$ exponential, $t$ monopole, PV, Regge\} regulators, respectively.
This supports the continued use of the $s$-dependent exponential regulator as was done in the JAM baseline model.

\subsection{Parton distributions in the pion}

\begin{figure}[t]
\centering
\includegraphics[width=0.83\linewidth]{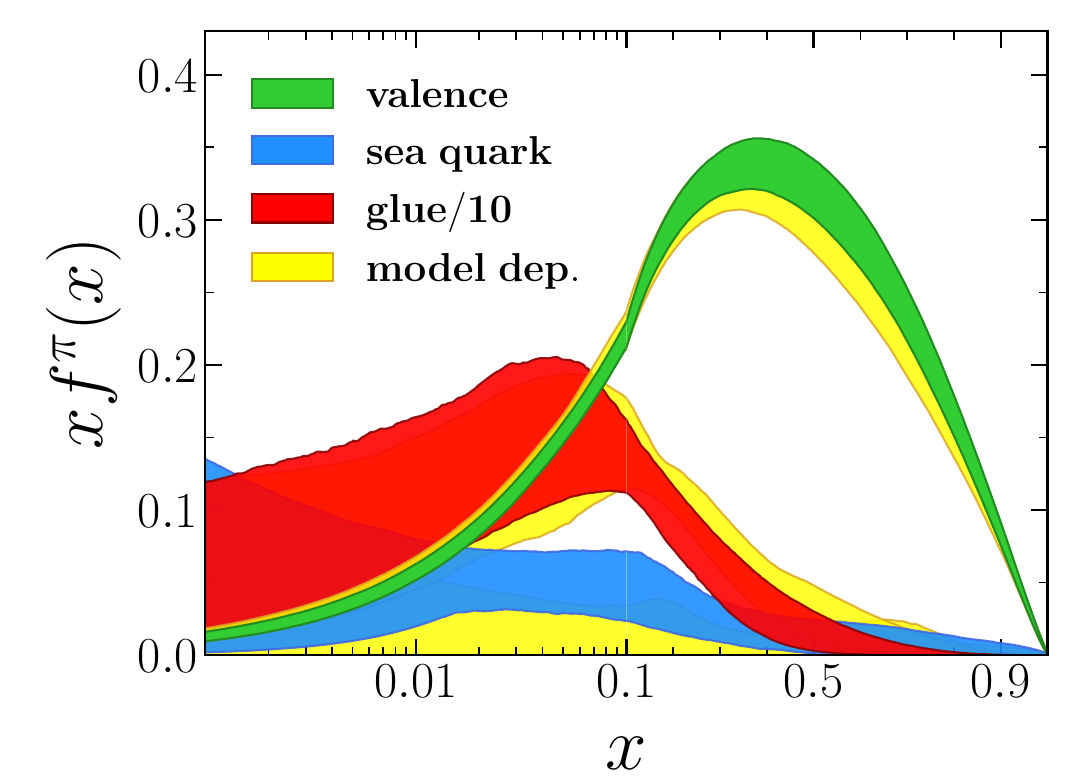}
\caption{Pion valence quark (green 68\% CI bands), sea quark (blue), and gluon scaled by 1/10 (red) PDFs versus $x$ at the input scale $\mu=m_c$ from the JAM extended data selection (see Table~\ref{tab:scenarios}). The uncertainty from the splitting function model is indicated by the yellow bands. \\}
\label{fig:final_pdfs}
\end{figure}

Having established the quality of the fits and description of the available datasets, we now present the final results for the pion PDFs in the JAM extended analysis.
In Fig.~\ref{fig:final_pdfs} we show the pion PDFs $x f^\pi$ for the valence, sea quark, and gluon (scaled by a factor 1/10), at the input scale $\mu=m_c$.
The main results use the $s$-dependent exponential regulator for the splitting function, and the systematic uncertainty from varying the splitting function model is relatively small compared with the other uncertainties.
While the results in Fig.~\ref{fig:final_pdfs} are obtained for the JAM extended dataset, including the additional H1 data (see Table~\ref{tab:scenarios}), they are effectively indistinguishable from those obtained for the JAM baseline dataset used in previous JAM analyses~\cite{Barry:2021osv}.

\begin{figure}[h]
\centering
\includegraphics[width=0.8\linewidth]{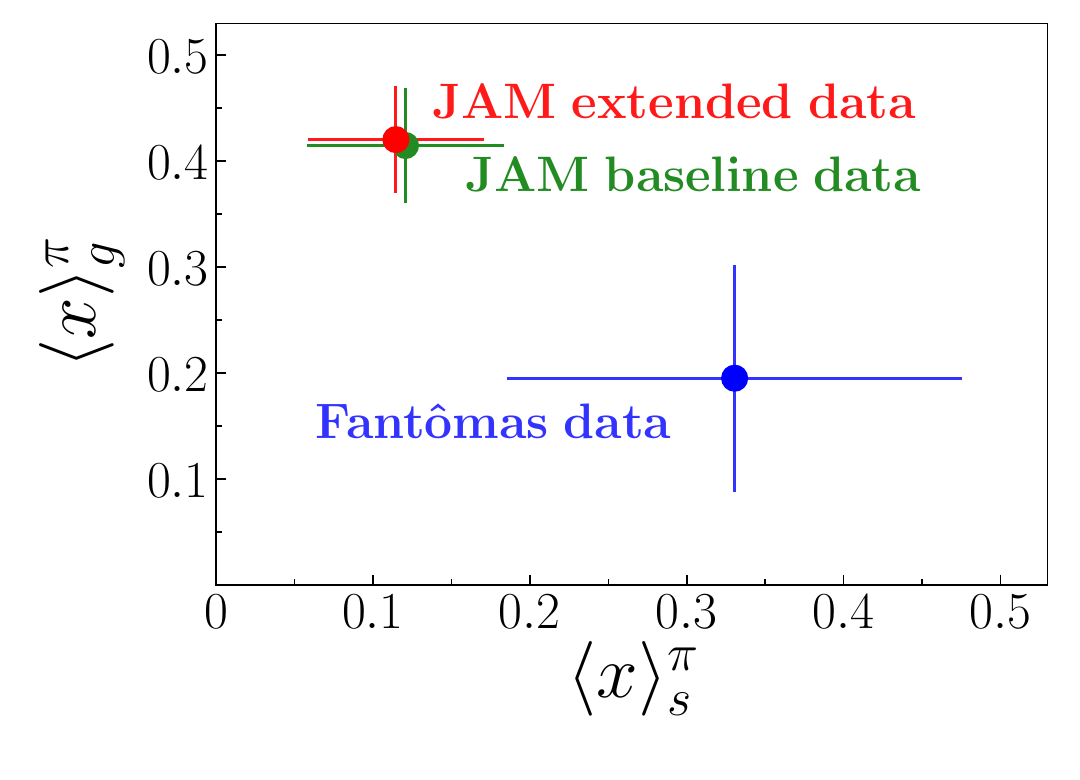}
\caption{Momentum fractions in the pion carried by gluons, $\langle x \rangle_g^\pi$, and sea quarks, $\langle x \rangle_s^\pi$, at the input scale $\mu=m_c$ from the JAM global analysis fitted to the baseline dataset (green), extended dataset (red), and the restricted H1 dataset used in the Fant\^omas analysis~\cite{Kotz:2023pbu} (blue).}
\label{fig:H1ext-moms}
\end{figure}

The similar results for the JAM baseline and extended dataset are also reflected in Fig.~\ref{fig:H1ext-moms}, where we show the average gluon and sea quark momentum fractions in the pion,
\begin{subequations}
\begin{eqnarray}
\langle x \rangle_g^\pi
&=& \int_0^1 \dd x\, x\, g^\pi(x,\mu),
\\
\langle x \rangle_s^\pi
&=& 6 \int_0^1 \dd x\, x\, q_s^\pi(x,\mu).
\end{eqnarray}
\end{subequations}
In particular, for the JAM extended results we find the values
$\langle x \rangle_g^\pi = 0.42(5)$
and
$\langle x \rangle_s^\pi = 0.12(6)$
at the input scale.
In contrast, the fits using the Fant\^omas data selection~\cite{Kotz:2023pbu} with the restricted H1 dataset give a comparatively smaller gluon momentum fraction, 
$\langle x \rangle_g^\pi = 0.20(11)$,
but a larger sea quark momentum fraction, 
$\langle x \rangle_s^\pi = 0.33(15)$, 
compared with the JAM result.

\section{Conclusion}
\label{sec:conclusion}

In this analysis we have performed a detailed reassessment of the determination of pion PDFs from pion-nucleus Drell-Yan and leading neutron electroproduction data, focusing on the origin of the differences between the JAM~\cite{Barry:2018ort, Cao:2021aci, Barry:2021osv, Barry:2022aix} and Fant\^{o}mas~\cite{Kotz:2023pbu} results for the valence, sea quark, and gluon distributions and their uncertainties.
In particular, by varying the LN data selection, the flexibility of the PDF parametrization, including positivity constraints, and the model for the pion--nucleon splitting function, we have identified which features of the extracted PDFs are driven by data and which arise from assumptions in the analysis.

For the analysis with the restricted H1 data selection, we found that the more flexible two-shape parametrization produces a large valence quark uncertainty at $x \sim 0.1$, similar to that observed in the Fant\^{o}mas analysis.
However, the replicas responsible for this broadening contained negative valence distributions and anomalously large sea-quark contributions at moderate and large $x$.
Imposing positivity removes these replicas and gives a valence distribution close to that obtained with the simpler, positive one-shape parametrization. 
The enlarged valence quark PDF uncertainty therefore appears to be a consequence of choosing a reduced dataset to constrain the PDFs, rather than a problem with parametrization flexibility of the previous JAM analyses.

The restricted H1 dataset also has insufficient leverage to distinguish between a fitted gluon PDF and one that vanishes at the input scale.
Once the full H1 and ZEUS LN coverage is included, the zero-gluon fit increases the total $\chi^2$ significantly.
The free-gluon fit reproduces the measured $x_\pi$ dependence more accurately across the H1 and ZEUS bins, while the zero-gluon result gives slopes that are inconsistent with the observed trends. 
Within the present theoretical framework, the combined LN data therefore clearly favor a nonzero gluon PDF in the pion at the input scale.

The extended analysis shows that the additional H1 measurements can be described within the one-pion exchange framework for $\bar{x}_L < 0.32$, whereas extending the ZEUS selection beyond $\bar{x}_L \approx 0.2$ leads to a poorer description. 
Among the splitting-function models considered, the $s$-dependent exponential regulator gives the best overall fit and remains stable when the additional H1 data are included, while other regulator forms require substantially larger cutoff masses to accommodate the higher-$\bar{x}_L$ data. 
Predictions based on the fitted splitting functions also describe well the independent $pp \to nX$ measurements at low transverse momentum, with the $s$-dependent exponential form again giving the best agreement.

The pion PDFs obtained with the extended dataset are consistent with the JAM baseline results, and their dependence on the splitting-function model is small compared with the PDF uncertainties.
In contrast, the restricted H1 selection allows a redistribution of momentum from gluons to sea quarks that is not favored once the broader LN coverage is included. 
The extracted gluon momentum fraction is sensitive to the lowest-$\bar{x}_L$ ZEUS bin and to the correlated systematic shifts required in its description. 
Comparisons~\cite{Chang:2026tgb} based only on the unshifted ZEUS central values inflate the tension with the JAM fit, in which the correlated shifts and normalization factors are penalized in the likelihood.
Removing the $\bar{x}_L=0.06$ data leads to some modification of the fitted splitting function and transfers momentum from the gluon to the sea; however, a zero input gluon remains strongly disfavored. 
This endpoint sensitivity contributes to the systematic uncertainty on the magnitude of the pion gluon distribution without weakening the evidence for a nonzero gluon.

More precise separation of the sea quark and gluon distributions should be possible with new tagged deep-inelastic measurements planned at Jefferson Lab~\cite{Montgomery:2017yua} and at future electron-ion colliders~\cite{Achenbach:2023pba, Anderle:2021wcy}.
Direct constraints on the pion PDFs will also come from pion-induced DY measurements at the COMPASS++ and AMBER facilities at CERN~\cite{Adams:2018pwt}.
Advances in lattice QCD calculations with higher statistics data, and reliable inclusion of flavor-singlet channels through signal improvement methods, will provide further constraining power complementary to experimental data.

\begin{acknowledgements}
This work was supported by the U.S. Department of Energy, Office of Science, Office of Nuclear Physics under contract no. 89243126CSC000213. The work of N.S. was supported by the DOE, Office of Science, Office of Nuclear Physics in the Early Career Program. The work of C.Y. was supported by the National Science Foundation, Research Experiences for Undergraduates Program, under Award No. 2348822. The work of P.B. was supported by the U.S.~Department of Energy, Office of Science, Office of Nuclear Physics, contract no.~DE-AC02-06CH11357, and the Scientific Discovery through Advanced Computing (SciDAC) award {\it Femtoscale Imaging of Nuclei using Exascale Platforms}.
\end{acknowledgements}

\bibliography{ref}
\end{document}